\documentclass{article}

\usepackage{PRIMEarxiv}
\usepackage{hyperref}       
\usepackage{url}            
\usepackage{booktabs}       
\usepackage{tabularx}       
\usepackage{longtable}
\usepackage{ltablex}        
\usepackage{amsfonts}       
\usepackage{amsmath}
\usepackage{nicefrac}       
\usepackage{microtype}      
\usepackage{lipsum}
\usepackage{fancyhdr}       
\usepackage{graphicx}       
\usepackage{siunitx} 
\usepackage{threeparttable}
\usepackage{ragged2e}
\usepackage{amsmath} 
\usepackage{textcomp} 
\usepackage{tipa}     
\usepackage{hyperref} 
\usepackage[numbers]{natbib}
\usepackage{float}
\usepackage[margin=1in]{geometry}
\usepackage[T1]{fontenc}
\usepackage{lmodern}

\DeclareUnicodeCharacter{02BE}{\textquoteright}

\graphicspath{{./supplement-figures/}}    

\newcommand{\beginsupplement}{%
        \setcounter{table}{0}
        \renewcommand{\thetable}{S\arabic{table}}%
        \setcounter{figure}{0}
        \renewcommand{\thefigure}{S\arabic{figure}}%
        \setcounter{section}{6}
        \renewcommand{\thesection}{S\arabic{section}}%
     }

\title{VisTopics: A Visual Semantic Unsupervised Approach to Topic Modeling of Video and Image Data
\thanks{\textit{\underline{Citation}}: 
\textbf{Lokmanoglu, A. D., \& Walter, D. (2025). \textit{VisTopics: A Visual Semantic Unsupervised Approach to Topic Modeling of Video and Image Data} (). arXiv. http://arxiv.org/}} 
}

\author{
  Ayse D. Lokmanoglu\thanks{Corresponding author. ORCID: \href{https://orcid.org/0000-0001-5049-7906}{0000-0001-5049-7906}} \\
  Emerging Media Studies, College of Communication \\
  Boston University, Boston, MA \\
  \texttt{alokman@bu.edu}
  \And
  Dror Walter\thanks{ORCID: \href{https://orcid.org/0000-0001-9870-998X}{0000-0001-9870-998X}} \\
  Department of Communication \\
  Georgia State University, Atlanta, GA \\
  \texttt{dwalter2@gsu.edu}
}

\begin{document}
\maketitle

\begin{abstract}
Understanding visual narratives is crucial for examining the evolving dynamics of media representation. This study introduces VisTopics, a computational framework designed to analyze large-scale visual datasets through an end-to-end pipeline encompassing frame extraction, deduplication, and semantic clustering. Applying VisTopics to a dataset of 452 NBC News videos resulted in reducing 11,070 frames to 6,928 deduplicated frames, which were then semantically analyzed to uncover 35 topics ranging from political events to environmental crises. By integrating Latent Dirichlet Allocation with caption-based semantic analysis, VisTopics demonstrates its potential to unravel patterns in visual framing across diverse contexts. This approach enables longitudinal studies and cross-platform comparisons, shedding light on the intersection of media, technology, and public discourse. The study validates the method’s reliability through human coding accuracy metrics and emphasizes its scalability for communication research. By bridging the gap between visual representation and semantic meaning, VisTopics provides a transformative tool for advancing the methodological toolkit in computational media studies. Future research may leverage VisTopics for comparative analyses across media outlets or geographic regions, offering insights into the shifting landscapes of media narratives and their societal implications.
\end{abstract}


\section{Introduction}

In recent years, computational social science has gained significant popularity in Communication, particularly advancements in natural language processing (NLP) and unsupervised methods in text analysis \cite{Maier2018Applying}. These methods have revolutionized how researchers study large volumes of textual data, enabling novel insights into political language, media framing, and public discourse \cite{Baden2020Hybrid, Cappella2017Vectors, Freelon2018Computational, Peng2024Automated, Pouwels2024Integrating, Walter2024Meta}. Despite this progress, a substantial gap persists in applying similar computational rigor to visual data, which has become increasingly central to contemporary communication landscapes \cite{Peng2024Automated, Pouwels2024Integrating}. This gap is especially pronounced given the rise of visually oriented platforms such as Instagram, TikTok, YouTube, and Rumble, which dominate social media usage \cite{Gottfried2024Americans}. According to a recent Pew Research Center report, nearly one-third of U.S. adults use video-based platforms, with Instagram—a visual-first platform—engaging almost half (47\%) of all adults \cite{Gottfried2024Americans}. Given this shift, it is crucial to recognize that visuals are now just as important as text for understanding social phenomena, particularly when analyzing the impact of media in digital spaces.

Visual communication, much like textual communication, is instrumental in shaping public opinion and social attitudes. Images in media, ranging from iconic photographs to social media visuals, have historically functioned as framing devices that influence how audiences perceive issues, events, and individuals \cite{Entman1991Framing, Grabe2009Image, Graber1990Seeing, Hariman2007No, Manor2018Visually}. Unlike text, visuals operate at the subtextual level, leveraging elements such as color, composition, and cultural symbolism to evoke emotional and cognitive responses \cite{Kress1996Reading, Messaris2001role}. They possess a unique ability to condense complex social phenomena into compelling narratives, as seen in their role in shaping attitudes toward war, political scandals, or social movements \cite{Geise2015Putting}. Visuals have also been distinguished from textual frames by their ability to add a stronger pathos to verbal messaging in news stories \cite{Brantner2011Effects, Graber1990Seeing, Parveen2020Visual}. Visuals, accordingly, have effectively shaped public attitudes toward visual subjects, including racial essentialism \cite{Said1981Covering}. In addition, images can be reproduced with high fidelity across time and space. This dual capacity for symbolic depth and emotional resonance underscores the need to expand computational approaches to include visual data in communication research.

One key advantage of visual data in social science research is its ability to convey complex messages more effectively than text \cite{Peng2024Automated, Pouwels2024Integrating}. In politically charged environments, images, such as those shared during online protests, are particularly powerful at mobilizing audiences by capturing attention and evoking emotional responses \cite{Casas2019Images}. Visual content not only draws more engagement on platforms like Instagram but also allows for strategic communication by politicians, who use these platforms to shape public perceptions and enhance their visibility \cite{Araujo2020Automated, Peng2021What, Peng2024Automated}. 

Researchers are often more interested in the meaning and context conveyed by content, which is why the focus tends to be on semantics \cite{Baden2020Hybrid, Jiang1997Semantic, Kroon2024Advancing, Walter2019News, Watanabe2021Latent, Xiong2016Semantic}. The relationship between visual and semantic similarity has been a key focus of numerous prior studies, reflecting the growing need to understand how visual data correlates with semantic meaning \cite{Brust2019Not, Groot2016stimulus, Deselaers2011Visual}. Semantics allow us to explore not just the surface-level attributes of visual data, but the deeper meanings, implications, and cultural symbols embedded within those visuals. This is crucial when analyzing how images, like those found in social media, advertising, or political campaigns, communicate messages to audiences. Deselaers \& Ferrari \cite{Deselaers2011Visual} developed a distance function that considers both semantic and visual similarity, demonstrating that this hybrid approach outperforms traditional distance functions that rely solely on visual characteristics.  

Recent developments in computational tools, particularly in unsupervised learning, deep neural networks, and Large Language Models, offer promising avenues for analyzing visual content at scale. These tools enable researchers to move beyond labor-intensive manual coding, allowing for the clustering and classification of large-scale visual datasets based on both semantic and visual properties \cite{Caron2018Deep, Chen2021Jigsaw}. However, despite these methodological advancements, most applications remain limited to static images, leaving video content underexplored. This oversight is significant, as video platforms increasingly serve as dominant spaces for political messaging, advocacy, and grassroots mobilization \cite{Casas2019Images, Lu2024Mobilizing}. 

Following the suggestions of Peng et al \cite{Peng2024Automated} in automated visual clustering, we propose a novel approach for image and video analysis that integrates unsupervised image clustering with large language models (LLMs) to analyze visual data. This end-to-end methodology covers every stage—from data extraction to the identification of thematic semantic structures within visual content—thereby providing a powerful framework for addressing theoretical questions in communication research, as well as giving the researcher the agency to choose their model of clustering. By leveraging the combined strengths of machine learning and NLP, this end-to-end methodology facilitates the identification of thematic and semantic structures within both static and dynamic visual content. This paper also extends current methods by incorporating video data, broadening the analytical scope and enabling the study of more complex and temporally rich datasets. In doing so, it addresses a critical gap in computational social science while offering a scalable framework for analyzing visual media's impact on public opinion and social behavior.

\section{Computational Methods for Visual Analysis}
Computational methods have revolutionized how social scientists engage with visual data, offering scalable solutions to processes that were once labor-intensive and limited in scope. By clustering images according to both visual and semantic properties, these methods have demonstrated immense potential for expanding our understanding of communication phenomena in digital spaces. Traditional content analysis approaches often rely on manual coding, which, while effective for small datasets, lacks the scalability required to analyze the vast quantities of visual data generated on platforms like Instagram, YouTube, and TikTok. Automated clustering techniques address this limitation by uncovering underlying patterns in visual data without the need for extensive manual intervention.

Before Zhang and Peng’s \cite{Zhang2022Image} work, several foundational studies explored unsupervised visual clustering, establishing methods that have shaped current approaches in computational social science. Traditional clustering techniques, like the Bag-of-Visual-Words model, which quantifies local visual elements like edges and textures—analogous to word frequencies in text analysis. While this model provided a baseline for image clustering, its inability to capture high-level features presented challenges for researchers analyzing datasets with complex semantic layers. To address these limitations, Caron et al. \cite{Caron2018Deep} introduced DeepCluster, a method that employs iterative clustering using k-means within feature spaces to generate pseudo-labels. These pseudo-labels progressively refine model layers, enabling the extraction of high-level features from unlabeled data. This iterative approach significantly improved the ability to identify cohesive semantic themes in visual datasets, particularly for social phenomena.

Building on these advancements, methods like Jigsaw Clustering \cite{Chen2021Jigsaw} and Clustervision \cite{Kwon2018Clustervision} have further enhanced the scalability and adaptability of unsupervised clustering techniques. Jigsaw Clustering uses a pretext task involving randomized image patches for spatial reconstruction, improving computational efficiency while capturing intricate visual representations. In contrast, Clustervision allows user-guided selection of high-quality clustering outcomes based on metrics such as the Calinski-Harabaz index and Silhouette coefficient, which enhance customization and interpretability. These innovations make unsupervised clustering particularly well-suited for complex datasets, such as social media imagery, where semantic distinctions are crucial for meaningful analysis.

Recent methodological advancements have incorporated self-supervised and transfer learning techniques to improve cluster accuracy and relevance. Self-supervised learning employs tasks like scrambling images to create meaningful distinctions within datasets, while transfer learning leverages pre-trained models such as ResNet to convert raw pixel data into formats suitable for computational analysis. Zhang and Peng \cite{Zhang2022Image} further advanced the field by focusing on low-dimensional representations that retain critical visual information. Their work emphasizes the importance of selecting effective feature extraction techniques, particularly in thematic domains such as protest imagery and climate change visuals, where nuanced interpretations are essential \cite{Lu2024Mobilizing, Peng2021What, Peng2023Agenda, Qian2024Convergence}.

These methods have proven particularly effective in identifying emergent patterns in dynamic social phenomena. For instance, Zhang and Peng \cite{Zhang2022Image} demonstrated how low-dimensional clustering could capture the fast-paced sharing of visual data during social movements. Traditional content analysis, which relies on human coders, would struggle to keep up with such rapid shifts, underscoring the importance of computational scalability.

While unsupervised clustering is valuable for exploring large-scale datasets, supervised learning methods offer precision in analyzing specific image features \cite{Courtois2023Computer, Lu2024Mobilizing, Peng2021What}. These approaches are especially useful for targeted analyses, such as identifying political branding elements or campaign visuals \cite{Courtois2023Computer}. By applying predefined labels to train models, supervised methods allow researchers to focus on specific research questions with a high degree of accuracy. Both supervised and unsupervised approaches rely on intermediate image representations derived through techniques like transfer learning, facilitating a deeper understanding of visual content across diverse datasets \cite{Courtois2023Computer, Lu2024Mobilizing}. 

The integration of convolutional neural networks (CNNs) into visual analysis workflows has further expanded the analytical capabilities of social scientists. CNNs excel in image classification tasks, enabling large-scale analyses of visual content with unprecedented accuracy \cite{Webb2020Images}. These advancements have unlocked new possibilities for understanding the role of visual media in shaping public opinion and driving social movements \cite{Zhang2022Image}.  As such, these methods have provided a significant leap forward in the automated analysis of large-scale visual datasets, opening up new possibilities for social science research.

OpenAI’s Contrastive Language-Image Pre-training (CLIP) \cite{Radford2021Learning} model represents a significant advancement, seamlessly aligning visual and textual modalities to enable large-scale semantic analysis. However, the model's architecture—a hallmark of black-box AI systems—poses challenges for interpretability and transparency. Furthermore, CLIP restricts researcher input by not allowing customization of the underlying model or flexibility in experimenting with prompts, which limits its adaptability to nuanced research contexts. This lack of insight and control raises concerns about reproducibility and potential biases. These limitations highlight the need for computational frameworks that prioritize explainability and user agency, ensuring that researchers can critically evaluate outputs and their implications.

For example, studies of climate change visuals across platforms like Twitter and Instagram have revealed that different content types—such as infographics versus nature landscapes—elicit varying levels of engagement, influencing how advocacy organizations tailor their messaging strategies \cite{Qian2024Convergence}. Similarly, analyses of social-mediated protests, including the Black Lives Matter movement, highlight the role of protest imagery in mobilizing public support during critical moments \cite{Lu2024Mobilizing}. These examples demonstrate the power of visual media in not only reflecting social realities but also actively shaping them.

\subsection{Challenges and Opportunities in Computational Analysis of Videos}
The increasing dominance of video content on digital platforms has introduced new complexities for computational social science \cite{Gottfried2024Americans, Nassauer2022Video}. Unlike static images, videos encompass temporal and dynamic dimensions, providing richer data but also presenting substantial analytical challenges. These challenges have limited the scalability of video analysis within communication research, leaving significant opportunities for methodological innovation.

\subsubsection{Challenges}
Analyzing video content requires addressing several unique challenges that arise from its complexity and structure. First, videos generate significantly larger datasets than static images, as they consist of sequential frames that must be processed and interpreted. This necessitates strategies for data reduction, such as frame sampling, to minimize redundancy while preserving meaningful information. Existing approaches, such as those proposed by Wang and Gupta \cite{Wang2015Unsupervised}, emphasize the importance of temporal consistency in extracting representative frames, yet balancing efficiency and accuracy remains an ongoing challenge.

Second, describing video content often involves combining visual and textual analysis, particularly when applying automated annotation methods. Video frames are inherently multidimensional, capturing not only visual features but also contextual cues that extend across time \cite{Nassauer2022Video}. Automated tools, such as image captioning models, offer a scalable solution but are constrained by limitations in accuracy and context sensitivity. These issues become particularly pronounced in datasets with complex semantic structures, such as political advertisements or protest footage.

Third, clustering methods traditionally designed for static images must be adapted to handle the sequential and multi-frame nature of video content. Techniques like Latent Dirichlet Allocation (LDA) and BERTopic, while effective for static data, may struggle to maintain coherence across frames. However, these methods demand significant computational resources and careful parameterization to ensure robust outputs.

\subsubsection{Recent Advances}
Recent advancements in computational methods have sought to address these challenges by leveraging unsupervised learning frameworks. Wang and Gupta \cite{Wang2015Unsupervised} introduced a method for learning visual representations from videos using temporal consistency as a supervisory signal. Their approach extracts patterns across frames by identifying shared features, enabling more effective clustering of video content.

Building on this foundation, Zhuang et al. \cite{Zhuang2020Unsupervised} proposed the use of deep neural embeddings to enhance unsupervised learning in video datasets. By embedding video frames into a shared feature space, their method enables researchers to uncover latent structures and thematic clusters, even in large-scale datasets. These advancements have paved the way for more nuanced analyses of video content, integrating both visual and semantic dimensions.

Videos provide an invaluable resource for understanding complex communication phenomena \cite{Nassauer2022Video}. In political communication, for instance, video-based campaign advertisements offer insights into the interplay between visual framing and narrative strategies \cite{Grabe2009Image}. Similarly, social movements increasingly rely on video content to mobilize audiences, with protest footage often serving as a catalyst for public engagement. By integrating clustering techniques with semantic analysis, researchers can explore these dynamics at scale, uncovering patterns that were previously inaccessible.

The integration of these methods into computational social science not only addresses the challenges of video analysis but also expands its theoretical and practical applications. By bridging the gap between static and dynamic visual data, video-based analyses open new avenues for studying communication processes in digital spaces.

Despite the integral role of images in human communication, visual communication remains underrepresented in media effects research \cite{Matthes2009What, Matthes2008Content, Peng2023Agenda, Walter2024Meta}. In a recent meta-analysis on framing, Walter and Ophir's \cite{Walter2024Meta} findings revealed that only 9.4\% of published papers addressed visual framing. Similarly, Peng et al. \cite{Peng2023Agenda} highlighted the slow growth of visual-related research in three flagship political communication journals, with fewer than 70 articles published in 2020. This limited focus underscores an ongoing need to analyze how visuals frame online public messaging.

Building on Peng’s foundational contributions to unsupervised image clustering \cite{Peng2021What, Peng2023Agenda, Peng2024Automated, Zhang2022Image}, this study seeks to couple unsupervised machine learning with AI-powered image captioning tools. Our aim is to introduce a package that enables semantic clustering for both images and videos, broadening the analytical toolkit available to communication researchers. We therefore introduce a method of semantic visual clustering in the sections that follow.

\section{Methods}
\subsection{Materials: NBC News Clips}
We chose NBC News clips for this study because of the network’s prominence as a mainstream news outlet and its influential role in shaping public opinion through visual and narrative strategies. In 2022, NBC's evening news program maintained an average audience of over 6.5 million viewers \cite{Pew2023Network}. As of the third quarter of 2024, NBCU News Group ranked as the \#1 digital news organization in the U.S., averaging 146 million monthly unique visitors, and on YouTube averaging 556 million monthly views \cite{NBCU2024NBCU}. News media have long been recognized as central actors in framing public discourse by not only selecting what topics to cover but also determining how these topics are presented \cite{Entman1991Framing, Geise2015Putting}. Visual framing, in particular, serves as a powerful tool for guiding audience interpretations, providing cues that reinforce specific perspectives or emotional responses \cite{Anden2008Abu, Brantner2011Effects}.

By focusing on NBC News, we aim to analyze how a major news organization utilizes visual content to contribute to framing processes and influence public understanding. NBC’s televised news clips offer a rich corpus that allows us to explore the interplay of visual and narrative elements in shaping audience perceptions. Previous research has demonstrated that mainstream news outlets rely heavily on visuals to frame topics such as political scandals, social movements, and public crises, often influencing broader societal discourses \cite{Anden2008Abu, Peng2023Agenda}. Our selection of NBC News ensures that our analysis is rooted in a context where these framing effects are likely to be both impactful and visible.

Televised news clips provide a unique opportunity for studying multimodal framing. Unlike static images or text, videos combine temporal and spatial dimensions, enabling us to examine how sequences of visuals and accompanying narratives work together to construct meaning. Through this multimodal lens, we can investigate how visual and verbal elements converge to evoke emotions, convey interpretations, and mobilize audiences \cite{Geise2015Putting, Grabe2001Explicating, Matthes2008Content}. NBC News’ use of high-quality visuals and its emphasis on storytelling make it an ideal choice for studying these phenomena.

Additionally, NBC News’ wide-reaching audience offers a robust context for understanding how visual framing influences public discourse at a national scale. By analyzing NBC’s clips, we aim to contribute to the growing body of research on visual framing in media effects and address the methodological challenges of working with large-scale video datasets. This research allows us to uncover how mainstream media outlets leverage visual content to frame events and shape public narratives in an increasingly digital media landscape.

\subsection{Analysis}
\begin{figure}
  \centering
  \includegraphics[width=0.5\textwidth]{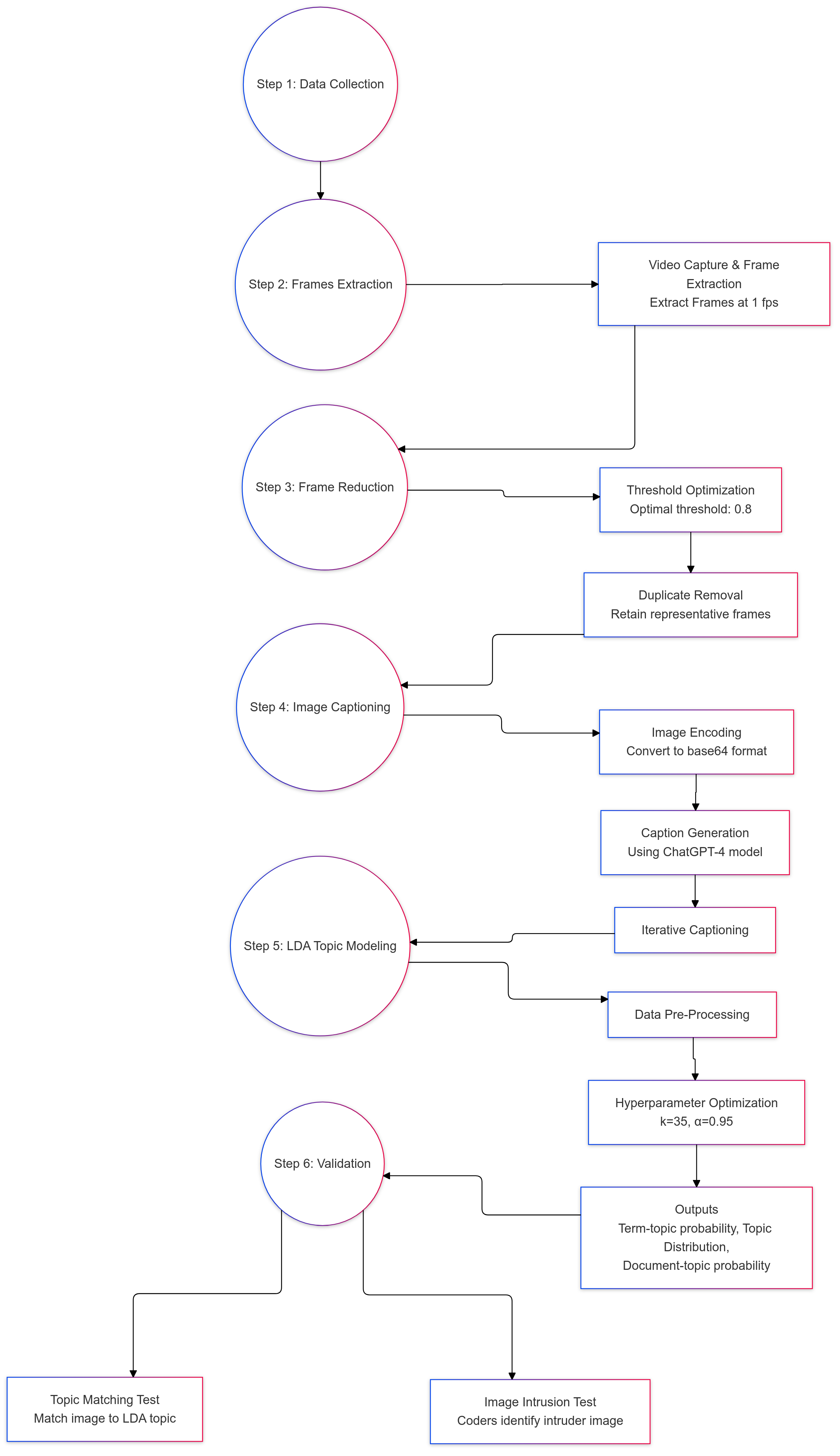} 
  \caption{Visual Clustering Workflow.}
  \label{fig:fig1}
\end{figure}

To develop a comprehensive pipeline for image clustering, we implemented a multi-step approach that spans from video file processing to the generation of semantic topics in a combined python package VisTopics  (see figure \ref{fig:fig1} for the visualization of the methods flow). This pipeline integrates Python 3.11.9 and R 4.4.0, offering flexibility for researchers to choose among clustering techniques such as Latent Dirichlet Allocation (LDA), BERTopic, or embedding-based methods. Computations were performed on the Shared Computing Cluster at [redacted]. Below, we detail each step of the process (for a full list of packages please see Supplement \ref{SupA}).

\subsubsection{Step 1: Data Collection}
The videos for analysis were collected from NBC News YouTube Channel on October 13, 2024. We downloaded the full playlist titled Top News from the channel, which included 2,300 videos at the time of download, starting at Nov 10, 2023, out of which 2,260 videos completed download successfully. To reduce analysis time, we sampled 20\% of the videos totaling in 452 videos used for analysis (total size 5.8 GB). 

\subsubsection{Step 2: Extracting Frames}
Video data is inherently dynamic, consisting of a sequence of static images called frames. Each frame represents a single, still image captured at a specific point in time within the video. Together, frames create the illusion of motion when played sequentially. For example, a video recorded at a frame rate of 30 frames per second (fps) contains 30 distinct images for each second of footage. These frames are the fundamental units of video analysis, offering a snapshot of visual content that can be analyzed independently or in context. 

To process our video dataset, we developed a custom Python pipeline that uses OpenCV \cite{opencv2024opencv} for video handling and frame extraction. Our goal was to extract representative frames from each video while maintaining computational efficiency. Rather than capturing every frame, which would result in an overwhelming amount of redundant data, we selected a frame rate of one frame per second (1 fps). This approach provided a manageable number of frames while retaining sufficient detail to capture the video's temporal dynamics. The main execution function, $extract_frames()$, iterates over the list of video files and extracts frames for each file, ensuring each video is processed sequentially and frames are stored in a structured directory system.

Within this function we had two tasks:
\begin{enumerate}
    \item Video Capture and Frame Extraction: The $cv2.VideoCapture()$ method from OpenCV is used to load the video. Frames are read sequentially in a while loop until the video file is exhausted. To enforce our desired frame rate of 1 fps, we calculated whether the current frame index matched our specified interval. This was determined by dividing the video’s frame rate by our target frame rate (1 frame per second). Frames that met this criterion were selected for extraction. For instance, if the original video was recorded at 30 fps, our script would extract every 30th frame to achieve the desired 1 fps rate.
    \item Frame Saving: Extracted frames are saved in JPEG format using cv2.imwrite() with names that reflect the frame sequence (e.g., frame\_1.jpg, frame\_2.jpg), supporting organized analysis and easy access for further processing.
\end{enumerate}

The extracted frames were organized into folders named after their corresponding video files, ensuring traceability and scalability for large datasets. By extracting one frame per second, we reduced the dataset's size without losing meaningful visual content. This frame sampling strategy also allowed us to balance computational efficiency with analytical rigor, ensuring that the extracted frames adequately represented the temporal and narrative complexity of the video data. During our analysis, the average processing time was as approximately ~0.1756 seconds (~175.6 milliseconds) per frame and ~4.3 seconds per video. In total, the pipeline processed 11,070 frames across 452 videos (Mean=24.491150, SD = 23.572832, Var = 555.67841), taking a total time of 32 minutes and 24.98 seconds to complete. 

\subsubsection{Step 3: Reducing Frames}
Given the substantial volume of frames extracted from the videos, reducing redundancy is an important step for computational efficiency and meaningful semantic analysis. To achieve this we utilized fastdup 2.9 \cite{visual2024visual} using a docker on shared cluster computing to identify and remove duplicate images efficiently. This tool leverages hash-based similarity detection to streamline the dataset while preserving its diversity. A key parameter in this process is the image hash threshold, which determines the degree of similarity between two images required for them to be considered duplicates. An image hash is a condensed digital fingerprint representing the content of an image. By encoding visual features such as textures, edges, and colors into a compact numerical representation, hashing enables efficient comparison between images. The image hash threshold specifies the maximum allowable difference between the hash values of two images to classify them as duplicates. Lower thresholds are more stringent, identifying only highly similar images, while higher thresholds allow for greater variation, potentially capturing near-duplicates.

\textbf{Threshold optimization}. In our pipeline, we conducted preliminary tests on a sample of 100 videos to identify an optimal threshold value. A threshold of 0.8 was chosen, striking a balance between minimizing redundancy and retaining representative diversity within the dataset. This value means that two images with hash similarities above 80\% were classified as duplicates. For instance, if a frame differed slightly in pixel alignment or lighting conditions but retained the same overall content, it was flagged as a duplicate under this threshold. This was calculated within each video, and not across videos. 

\textbf{Duplicate identification and removal}. Following the duplicate identification, the function iterated through subfolders corresponding to individual videos, detecting duplicate images and cataloging them alongside their original counterparts. Only the most representative frame from each duplicate set was retained for analysis. The final list of duplicate frames was compiled in a table format, recording original frames and duplicates in a for further analysis.

By employing this threshold-driven approach, we significantly reduced the size of the dataset while preserving its informational integrity, our dataset was reduced from 11,070 frames to 6,928 frames (716M size), with an average 15.33 per video (SD= 12.522577, Var= 156.81494). The whole process took 01:18:12 for 452 videos, averaging ~10.38 seconds per video. The resulting streamlined dataset was better suited for semantic analysis and clustering, enabling more efficient and focused computational processing. 

\subsubsection{Step 4: Image Captioning}
The emergence of large language models (LLMs) offer new ways to analyze both text and visuals \cite{Gilardi2023ChatGPT, Hoes2023Leveraging, Vreese2023AI, Wu2023Large}. Tools like ChatGPT have proven effective for annotation tasks, achieving higher accuracy and intercoder agreement than traditional methods \cite{Gilardi2023ChatGPT}. While LLMs have shown limitations in handling ambiguous or context-dependent content \cite{Hoes2023Leveraging}, their ability to scale analyses across diverse datasets marks a significant step forward for computational social science, thus we selected to use OpenAI model ChatGpt4-o \cite{OpenAI2024ChatGPT} to caption images. Our function contains 3 steps:
\begin{enumerate}
    \item Image Encoding: To prepare each image for captioning, our function reads and encodes the image file in base64 format. This step is critical as the OpenAI API expects images in base64 to process within its chat model.
    \item Caption Generation: Our function then sends a request to OpenAI’s API to generate an image caption. It uses a prompt instructing the model to provide a concise description of the image scene or characters. The response, if successful, returns a caption describing the scene, which is then stored. We tested different prompts on a sample on different types of video frames as well (political ad campaigns, Eurovision music videos), and found the following prompt to be most effective in capturing the semantic meaning of the frame:
    \begin{enumerate}
        \item “Directly describe with brevity and as brief as possible the scene or characters without any introductory phrase like 'This image shows', 'In the scene', 'This image depicts' or similar phrases. If there is a text in the image mention there is a text but do not caption the text, just start describing the scene please. If you recognize historical figures and current celebrities and politicians in the picture give their full name, but don't give the whole background about who they are.”
        \item This approach yielded captions that were precise and relevant to the visual content.
    \end{enumerate}
    \item Iterative Captioning: The function processes all images in a specified directory by encoding each and retrieving captions sequentially. A random delay is implemented between requests to prevent server overload. Captions are stored in real-time in a CSV file, with the path and generated caption recorded for each image.
\end{enumerate}

This method ensures efficient handling of potentially large image datasets, supports real-time data capture in the CSV, and maintains processing continuity by logging each step of the captioning workflow. The total time taken for 6,928 images was 10 hours, 34 minutes, and 1.71 seconds, averaging ~5.49 seconds per frame (including the 1 to 5 second random delay between each frames), costing total of $\$10.88$. 

\subsubsection{Step 5: LDA}
To uncover latent thematic structures within our dataset of image captions, we employed Latent Dirichlet Allocation (LDA), a widely used unsupervised machine learning technique for computational content analysis. LDA has been extensively applied in social science research to analyze expansive text corpora while mitigating researcher bias \cite{Del2016spreading, Walter2019News}. As a generative probabilistic model, LDA identifies latent topics by modeling the relationships between words within documents and across the corpus, offering researchers a means of inductively detecting thematic patterns based on term co-occurrence  \cite[pp.~93--94]{Maier2018Applying}.

We followed established pre-processing guidelines to prepare our image captions for LDA analysis, adapting strategies commonly used for social media text datasets \cite{Jacobi2016Quantitative, Maier2018Applying}. First, we conducted data cleaning by removing non-English captions, which helped maintain linguistic uniformity across the dataset (there was none in our dataset). Punctuation, numbers, and special characters, including HTML tags and non-standard symbols, were stripped from the text. Additionally, captions with fewer than 10 characters were excluded to eliminate entries that contained insufficient information.

To address duplicate handling, we identified and removed all repeated captions to prevent their disproportionate influence on the topic modeling process. These duplicates were later reintroduced after the LDA modeling phase to ensure their accurate representation in terms of theme prevalence and volume distribution.

Tokenization and filtering further refined the dataset. Captions were tokenized into individual words, and we removed tokens that were shorter than three characters to reduce noise from textual artifacts. Words that appeared in more than 50\% or fewer than 0.005\% of the captions (approximately 10 instances) were also excluded, focusing the analysis on meaningful patterns. We also removed stopwords to ensure that the resulting topics were not influenced by common, low-informational terms. This included a combination of standard English stopwords, domain-specific terms (e.g., news, NBC), and visually descriptive elements (e.g., colors) that could skew the results. Stemming was deliberately avoided, as previous research has shown that stemming can negatively impact the coherence of topics generated by LDA models \cite{Schofield2016Comparing}.

Lastly, these pre-processing steps ensured that the dataset was streamlined and optimized for the topic modeling process, enabling the LDA algorithm to effectively uncover latent semantic patterns and themes within the captions. This careful preparation was essential for producing interpretable and reliable results from the analysis.

For the Latent Dirichlet Allocation (LDA) analysis, we optimized two key hyperparameters, $k$ (the number of topics) and $\alpha$ (topic sparsity), in accordance with best practices from prior research \cite{Jacobi2016Quantitative}. To identify the most interpretable structure for the data, we tested $k$ values ranging from 5 to 100, examining models with varying numbers of topics to evaluate their coherence and applicability. Additionally, we adjusted $\alpha$ levels within a range of $25\/k$ to $1\/k$ to control the tendency for documents to exhibit multiple topics. Using 5-fold cross-validation and perplexity as our evaluation metrics, we determined that the optimal model consisted of 35 topics with an $\alpha$ value of $2\/35$ ($\sim 0.06$). This configuration provided a balance between interpretability and coherence, ensuring that the topics accurately reflected meaningful structures within the dataset.

The final LDA model yielded three critical outputs, offering a comprehensive understanding of the data. First, the term-topic probability ($\beta$) provided the probability distribution of terms within each topic, enabling us to identify the defining terms for each cluster. Second, the topic distribution across the corpus ($\theta$) highlighted the prevalence of each topic within the entire dataset, offering insights into the dominant themes. Finally, the document-topic probability ($\gamma$) indicated the likelihood of each topic being associated with individual captions, facilitating the clustering of similar captions into coherent groups.

To account for previously excluded duplicate captions, we reintroduced them during the post-modeling analysis phase. Each duplicate's topic distribution was assigned based on the most similar original captions, ensuring their representation within the final dataset and preserving the integrity of the thematic patterns. By applying LDA to our dataset of image captions, we identified latent semantic themes that inform the clustering process. These insights provided the foundation for further analysis, allowing us to systematically explore patterns and thematic structures within the visual content.

\subsubsection{Step 6: Validation}
To validate the coherence and interpretability of the LDA-generated topics, we employed a human evaluation process involving two coders trained on a separate dataset of Eurovision-related images. This external dataset ensured that coders were familiar with the principles of image categorization and clustering but unbiased toward the specific data and topics in our study. The validation consisted of two structured tasks designed to assess topic quality and alignment between visual and semantic themes. The tasks were based on the commonly used methods of word intrusion and topic intrusion \cite{Chang2009Reading}.

\textbf{Task 1: Image Intrusion Test}. The first validation test was based on word intrusion tests commonly used with topic modeling \cite{Chang2009Reading}, which we titled Image Intrusion Test. For this validation test coders were provided with a set of six images. Each set was constructed using five images strongly associated with a single LDA-generated topic and one "intruder" image from a different random topic. Coders were asked to identify the image that did not belong within the set of six images. This task assessed the internal coherence of each topic by evaluating whether the coders could reliably identify the intruding image. A high agreement rate among coders indicated that the LDA topics effectively grouped semantically and visually similar images.

\textbf{Task 2: Topic Matching Test}. The second test, which we titled the Topic Matching Test, evaluated the alignment of individual images with LDA topics. Coders were presented with four rows of four images. Each row included four images strongly associated with a single LDA-generated topic (for a total of four topics). Coders were then presented with an image drawn from one of these four topics and were tasked with matching it to the correct row. This test assessed whether the semantic and visual themes identified by the LDA model were intuitive and easily distinguishable. The accuracy of the coders in correctly matching images to topics provided an additional measure of topic validity.

Both validation tasks consisted of 105 items each (totaling in 210 items for each coder). Coders completed the tasks independently, with instructions emphasizing quick but thoughtful responses to simulate a realistic evaluation environment. The process was designed to be engaging and efficient, encouraging coders to rely on their intuitive understanding of visual and semantic groupings. Inter-coder agreement was calculated for both tasks, providing a quantitative measure of the reliability of the LDA topics.

By incorporating these human validation tests, we ensured that the topics generated by the LDA model were not only statistically robust but also meaningful and interpretable from a human perspective. 

\section{Results}
Table \ref{tab:table1} 1 outlines the computational pipeline's efficiency and scalability for processing and analyzing visual data. The collection of 2,300 NBC News videos was completed in less than an hour, with a total dataset size of 29.6 GB, later reduced to 452 images with a total size of 5.8 GB. Frame extraction yielded 11,070 frames in 32 minutes and 25 seconds, with an average of 4.30 seconds per video and 0.175 seconds per frame, producing a dataset of 1.2 GB. Using image-hash similarity reduction, the dataset was refined to 6,928 unique frames over 1 hour, 18 minutes, and 12 seconds, averaging 10.38 seconds per video and 0.63 seconds per frame, resulting in a reduced size of 716 MB. Captioning these frames with OpenAI tools required over 10 hours, at a cost of $\$10.88$, with an average processing time of 5.49 seconds per frame. Altogether, the process of preparing the visual dataset for clustering took 12 hours, 33 minutes, and 3 seconds.

\begin{table}[htbp]
  \centering
  \caption{Pipeline Resource Metrics Across Processing Stages} \label{tab:table1}
  \begin{tabularx}{\textwidth}{lccccc}
    \toprule
    \textbf{Stage} & 
    \textbf{Total Time} & 
    \textbf{N (Videos or Frames)} & 
    \textbf{Avg Time/Video (s)} & 
    \textbf{Avg Time/Frame (s)} & 
    \textbf{Total Size} \\
    \midrule
    Data Collection & 00:08:24 & 452 Videos & 1.12 & -- & 5.8 GB \\
    Frame Extraction & 00:32:25 & 11,070 Frames & 4.30 & 0.175 & 1.2 GB \\
    Frame Reduction (Deduplication) & 01:18:12 & 6,928 Frames & 10.38 & 0.63 & 716 MB \\
    Image Captioning & 10:34:02 & 6,928 Frames & -- & 5.49 & -- \\
    LDA Analysis & -- & 6,928 Frames & -- & -- & -- \\
    \bottomrule
  \end{tabularx}
\end{table}

The analysis of the final dataset of 6,928 frames revealed 35 distinct semantic topics, representing the thematic breadth of NBC News’ visual narratives. These topics, identified using Latent Dirichlet Allocation (LDA), spanned political events, environmental crises, and human-interest stories. Table \ref{tab:table2} provides an overview of the top words and representative captions for each topic, with additional details (and full image clusters) available in Supplement \ref{SubB}, figures \ref{fig:figb1} to \ref{fig:figb35}.

\begin{longtable}{p{0.7cm}p{4.8cm}p{9cm}}
\caption{Top Words and Captions from LDA Model ($k=35$)}\label{tab:table2}\\
\toprule
\textbf{Topic} & \textbf{Top Words (N=10)} & \textbf{Top Captions (N=3)} \\
\midrule
\endfirsthead

\multicolumn{3}{c}%
{{\bfseries \tablename\ \thetable{} -- continued from previous page}} \\
\toprule
\textbf{Topic} & \textbf{Top Words (N=10)} & \textbf{Top Captions (N=3)} \\
\midrule
\endhead
\bottomrule
\endfoot
X1 & police; officers; lights; vehicles; emergency; scene; vehicle; building; street; shooting & Urban area with a street, cars, and a fence. Signs for Rainbow Bridge to U.S.A. exist. News headline about a car explosion at the U.S.-Canada border.; Police officers in riot gear moving towards makeshift wooden structures and tents with metal barricades in Los Angeles. Text overlay present.; Bridge with signs pointing towards the USA and Canada. Text overlay about an FBI investigation related to a vehicle explosion near Niagara Falls. \\
X2 & military; soldiers; gear; uniform; personnel; helicopter; soldier; camouflage; syria; iraq & Sgt. Breonna Moffett in military uniform with text "Fallen American Soldiers Return Home" in a news banner.; A person wearing camouflage and a helmet is inside a military vehicle, surrounded by equipment. Text at the bottom of the image.; Helicopter rescue operation over rocky terrain, rescue personnel guiding a stretcher. Text present. \\
X3 & trump; donald; speaking; american; flags; harris; podium; stage; debate; microphone & Poll results comparing enthusiasm levels among supporters of Donald Trump, Ron DeSantis, and Nikki Haley with corresponding percentages.; Donald Trump standing at a podium with outstretched arms, background features American flags and campaign text, television news ticker at the bottom.; Donald Trump looking down at something in a suit, with overlapping images of a crowd and American flag. Text on the bottom reads "Iowa caucuses now just one week out." \\
X4 & israel; netanyahu; israeli; benjamin; gaza; hamas; hostages; war; fire; wall & Skyline of Jerusalem, featuring the Dome of the Rock with text overlay related to the Israel-Hamas conflict.; Images of missing individuals with "KIDNAPPED" labels, names, ages, and request for help. News channel banner about Israel-Hamas war hostages.; Flyers with images attached to a railing, accompanied by a text overlay about long-term support for Israel-Hamas war hostages. \\
X5 & food; fast; post; discussing; voting; response; pay; voice; counter; media & Combo meal menu display featuring a Big Mac, fries, and Coca-Cola; text present. News banner questioning fast food pay raise effects on prices.; Gretchen Whitmer, Josh Shapiro, Gavin Newsom, Roy Cooper, Andy Beshear, JB Pritzker, and Mark Kelly are displayed with their respective states. There is text.; Eight portraits with names and titles set against a green field; includes Gretchen Whitmer, Tim Walz, Roy Cooper, Josh Shapiro, Mark Kelly, JB Pritzker, Andy Beshear, and Pete Buttigieg. News banner at the bottom. \\
X6 & child; holding; woman; smiling; young; children; kate; princess; baby; adult & A magazine spread with a family photo of Catherine, Princess of Wales, with three children in a garden. Texts on the page discuss the family picture. Breaking news ticker: "PRINCESS KATE GIVES CANCER FIGHT UPDATE."; Document on a wooden surface titled "WanaBana" detailing a recall of apple cinnamon fruit purée pouches due to elevated lead levels. Breaking news banner below mentions FDA investigation into toxic applesauce after children fall ill.; Package of Wana Bana Apple Cinnamon Fruit Puree. News headline about FDA investigating toxic applesauce after kids fall ill. \\
X7 & person; holding; hand; corner; george; santos; wearing; phone; wooden; nearby & Person reclining on a bench with a neck pillow, using a phone, feet propped on a suitcase; pink water bottle nearby. Text present.; A person holding a phone with a charger attached, wearing jeans. Text about new laws in 2024 is present.; Hand stubbing out a cigarette butt in a cannabis leaf-shaped ashtray on a wooden surface, with WBAL and NBC logos. \\
X8 & hair; woman; wearing; man; speaking; glasses; shirt; person; long; short & Woman with long blonde hair and glasses, wearing a denim shirt, speaking outdoors. Text overlay regarding NBC News Investigates.; A woman with short, wavy blonde hair is smiling, wearing a teal blouse, sitting indoors with a beige background.; Man with a beard and mustache in a suit appears on a screen. Text at the bottom. Room with artwork, candle, and glass cabinet in the background. \\
X9 & view; people; aerial; building; trees; large; area; surrounded; vehicles; cars & Aerial view of a desert landscape with a long border wall. Groups of people and several vehicles are gathered along the wall. Desert terrain with sparse vegetation and distant hills.; Aerial view of a school campus with a football field, parking lot, and surrounding buildings. There is visible text in the image.; Aerial view of a dam or bridge over a river surrounded by green fields and trees, with a road crossing it. Text overlay and news banner at the bottom. \\
X10 & field; football; star; crash; paris; university; usa; track; stadium; wearing & Hockey player in a Columbus Blue Jackets jersey with the name "Gaudreau" and the number 13 skating on ice. Text present about a statement from the Columbus Blue Jackets. News ticker reads "NHL star and brother killed in crash."; Two ice hockey players wearing jerseys with the name "Gaudreau" and numbers 21 and 13, kneeling on the ice. Lower text banner about a crash involving an NHL star and brother.; Football player in a blue and yellow uniform, number 9, throwing a football. Stands in front of a crowd; text present about a sideline event regarding the Michigan head coach. \\
X11 & wearing; man; outdoors; shirt; standing; person; jacket; woman; stands; cap & Kari Lake speaking into a microphone, wearing a red outfit. News reporter standing outdoors, wearing a jacket. Text present overlaid on the image.; Kevin McCarthy speaking outdoors to a seated audience, wearing a baseball cap and holding a microphone. Text present in the image.; A young couple is standing together, smiling. The man wears a white naval uniform, and the woman wears a dress with a wide-brimmed hat. There is a floral backdrop with candles. \\
X12 & flag; american; podium; speaking; flags; microphones; man; emblem; standing; house & Alejandro Mayorkas speaking at a podium, with Merrick Garland in the background. Seal of the Department of Justice visible. Text: "Breaking News: House Votes to Impeach Mayorkas."; Lloyd Austin speaking at a podium during the IISS Shangri-La Dialogue event, flanked by teleprompters, with text overlay regarding Pentagon and White House communication.; Ayatollah Ali Khamenei speaking at a podium with multiple microphones. Image of Ebrahim Raisi superimposed. Background with the Iranian flag. News ticker at the bottom. \\
X13 & airplane; medical; hospital; equipment; interior; bed; stairs; aircraft; flight; boeing & A person lies in a hospital bed wearing an oxygen mask, with a medical professional in gloves attending. Medical equipment is nearby. Text is present in the image.; Person receiving medical attention, lying on the floor with oxygen mask, surrounded by medical personnel, with a towel underneath. Text present.; Airplane interior with a missing window, exposed insulation, and passenger seats visible; airport tarmac and vehicles outside. Text present. \\
X14 & flooded; partially; water; trees; submerged; street; area; flooding; emergency; flood & Flooded town with submerged streets and buildings, a bridge crossing over brown, churning water, and trees partially underwater.; Flooded street with rescue workers and residents wading through water, assisting each other; vehicles partially submerged in the background. Text present.; Floodwaters rushing through a forested area, with trees and debris in the water. Text overlay about historic flooding emergency in the Midwest. \\
X15 & suit; man; wearing; tie; speaking; microphone; backdrop; press; speaks; glasses & Bernie Sanders speaking in front of a Capitol building backdrop, wearing a suit and glasses, with "Meet the Press" logo visible.; Rudy Giuliani and a man sitting together at a table; one in a suit and tie, the other in a pink shirt. News text present.; Ebrahim Raisi speaking at a podium with microphones, wearing a black turban and glasses. Text present in the image. \\
X16 & large; screen; displaying; screens; display; standing; people; setting; images; anchor & Three people talking outside Radio City Music Hall with a television broadcast overlay featuring an eclipse and totality timings.; Three large blue and white screens displaying "DNC 2024" with red stripes, reflecting off a shiny floor. Text below about the nomination battle and DNC timing.; News anchor in a studio with "Hallie Jackson NOW" displayed on the screen. Breaking news banner below about a shooting on the University of Nevada, Las Vegas campus. \\
X17 & building; corner; screen; colorful; power; design; displaying; abstract; lines; featuring & Smartphone on a blue background, FBI emblem, "Consumer Alert" and "Massive AT\&T Customer Data Breach" text at the bottom. NBC News logo in the corner.; Smartphone displaying TikTok logo on laptop keyboard; news banner about House passing bill to ban TikTok.; Smartphone screen with blue digital code, text about phone numbers compromising personal relationships, "Massive AT\&T Customer Data Breach" headline, Consumer Alert banner. \\
X18 & mosquito; sign; hand; store; holding; viruses; federal; hands; setting; interest & Board of Governors of the Federal Reserve System emblem with text about June 2022 inflation at 9.1\%. News ticker reads "Federal Reserve cuts interest rates." Background of U.S. currency.; Gallon jugs of 365 Whole Milk on a store shelf; news ticker about health concerns related to bird flu virus in pasteurized milk.; Lab equipment and containers with a gloved hand inside a laboratory setting; a news banner reads "Breaking News" and "Senate Showdown Over IVF Protections" with NBC News logo. \\
X19 & sky; smoke; buildings; large; trees; flames; cloudy; fire; area; landscape & Wildfire with intense red and orange flames in the sky; silhouetted trees in the foreground. Text about dangerous heat and wildfires on the West Coast.; Thermometer graphic with 14 degrees, cloudy sky background, text indicating dangerous heat and wildfires on West Coast.; Vibrant purple sky with Northern Lights over a backyard pool, trees silhouetted against the horizon; news ticker at the bottom. \\
X20 & report; trial; hunter; special; indicating; biden; guilty; tax; labeled; documents & Legal documents with text "Guilty of 3 Felony Counts" highlighted, mentioning Robert Hunter Biden. News ticker states "Hunter Biden guilty on all counts in gun trial."; A collection of money, documents, and electronic devices; overlay includes text indicating a guilty verdict for Sen. Menendez.; U.S. Secret Service badge labeled "Sergeant" on a dark background with text at the bottom regarding a Secret Service agent related to Biden's granddaughter. \\
X21 & screen; desk; suit; split; anchor; man; anchors; wearing; speaking; broadcast & Two female news anchors at a desk with a cityscape backdrop; inset with a man standing in front of screens labeled with election graphics; another inset shows a male reporter speaking outdoors in a snowy setting. Text is present.; Two news anchors sit at a desk in front of a city skyline backdrop, one wearing a yellow jacket and glasses, the other in a dark suit. A woman in a red top appears on a video call to the right. Text is present on the screen.; Two news anchors sitting at a desk with papers, wearing red outfits, in front of a city backdrop; another woman appearing on a screen to the right in a newsroom, wearing a red sweater. Text banner at the bottom. \\
X22 & trump; donald; courtroom; sketch; wearing; america; setting; trial; speaking; make & Donald Trump standing in a courtroom, lawyer speaking, judge seated behind bench, others present, with courtroom setting. News text overlay present.; Courtroom setting with empty seats, polished banister, and text overlay about "Trump testifies in civil defamation trial."; Donald Trump speaks while Hope Hicks listens intently. Text about Hope Hicks testifying in a trial at the bottom. \\
X23 & bridge; water; ship; ocean; sky; containers; collapsed; large; cargo; boat & Aerial view of a busy port with numerous shipping containers stacked and large cranes; adjacent waterway with a docked ship. Text about a major strike at U.S. ports.; Aerial view of a U.S. Army Central floating pier at sea, featuring military equipment and personnel. Ocean waves surround the structure. Text present: "Gaza Aid Setbacks on U.S.-Built Pier."; Israeli navy personnel on patrol boat off Gaza coast, clear sky, distant cargo ship. Text present. \\
X24 & biden; joe; podium; speaking; american; flags; harris; flag; kamala; standing & Kamala Harris and Pete Buttigieg superimposed images; Harris in front of flags, Buttigieg speaking at podium with seal, overlay text about election and NBC logo.; Joe Biden at a podium with the presidential seal, flanked by American flags, in front of a fireplace and painting, delivering a speech. Text present.; Joe Biden on stage, gesturing with a smile, flanked by American flags, podium and teleprompter visible, text overlay at the bottom. \\
X25 & lit; people; dimly; building; walking; scene; night; bright; inside; person & School hallway with colorful murals, silhouettes of children on walls, polished floors, and dim lighting. Overlay of legs walking. NBC Nightly News logo present.; Crowded outdoor market with numerous people, stalls filled with various items, and stacks of packaged goods. Text present.; Four masked individuals in hoodies moving stealthily inside a room, caught on a security camera at night. Text is present in the image. \\
X26 & holding; people; crowd; signs; surrounded; flags; large; gathered; protest; protesters & People holding Palestinian flags, protest scene near a monument and lamppost; text about political pressure for ceasefire.; Mega Millions lottery tickets with NBC News NOW text overlay, mentioning a winning ticket sold in New Jersey worth \$1.13 billion.; Hands holding a Mega Millions lottery ticket. Text overlay notes a \$1.13 billion win in New Jersey. \\
X27 & fence; wire; barbed; spacecraft; space; rocket; border; nasa; earth; launch & Rocket launch pad with Boeing Starliner spacecraft, NASA logo, text overlay about the historic mission.; Blue tarp and chain-link fence; news banner below reads about a suspected serial killer captured in L.A.; Spacecraft in orbit above Earth, with partial view of planet below. NASA logo visible. Text present. \\
X28 & snow; road; covered; trees; street; winter; highway; traffic; person; scene & Bus dashboard view with hands on the steering wheel, following a tow truck on a snowy road. Trees line the road; snowy landscape. Text indicating news about tornadoes and winter storm.; Cars traveling on a snow-covered highway beneath an overpass, overcast sky, road signs visible, and trees lining the roadside. Text present at the bottom.; Snowy street with reduced visibility, vehicles in motion, person walking in deep snow at the top, multiple frames, text present. \\
X29 & people; group; men; wearing; surrounded; walking; attire; standing; suits; setting & Audience members clapping, some wearing hats and badges, at an event. Bright clothing and lanyards noticeable. Text in lower right corner.; People standing in line with umbrellas, fans, and hats to shield from the sun, surrounded by metal barriers and greenery. Casual summer attire, several individuals using handheld fans.; Group of people standing in a line, participating in a prayer or solemn gathering, wearing various clothing including traditional attire. Nighttime setting with bright lighting. Text present in the image. \\
X30 & map; states; weather; united; highlighting; indicating; showing; areas; severe; highlighted & Weather map of the West Coast of the United States, highlighting areas including Medford, Eureka, Chico, San Francisco, Bakersfield, Los Angeles, and San Diego, showing rainfall forecast with color-coded intensity.; Map of Europe with a storm system marked in northern Europe. Cities including London, Paris, Berlin, Vienna, and Marseille are labeled. Weather graphics and fronts are highlighted above the map. Text regarding weather information is present.; Weather map with colored regions indicating rain, ice, and snow across the western United States. Cities like Seattle, Portland, and San Francisco are labeled. Text on the left describes expected weather conditions. \\
X31 & hurricane; milton; map; weather; florida; showing; storm; tampa; large; stormy & Weather map tracking Hurricane Milton, Category 4, with swirling bands over Florida, details on wind speeds and movement.; Person in a blue rain jacket reporting with a sandbag being filled; weather map showing hurricane path. Text indicates breaking news about Hurricane Francine.; Palm trees swaying in strong wind; text overlay indicates hurricane speed, with map and "Breaking News" banner about Hurricane Helene targeting Florida. \\
X32 & people; sitting; seated; table; room; screen; setting; engaged; chairs; conversation & U.S. Senate chamber with senators seated at desks on a blue carpet, an official at a central desk, staff members present, flags against the walls, and ornate decorations. Text present.; Two women sitting at a table in an office or library setting, looking at a document. There are framed pictures on the wall behind them. Text overlay present.; Two people engaged in conversation indoors by a stained glass window, standing under a decorative lamp. Text present. \\
X33 & court; supreme; abortion; building; decision; texas; states; document; media; law & Document titled "Supreme Court of Alabama" with details of a case involving James LePage and Emily LePage. Overlay text reads "Breaking News" with a headline about Alabama clinics pausing IVF after a ruling.; The United States Supreme Court building with a breaking news banner stating the court acknowledges a document in a pending abortion case was accidentally posted online.; NBC News headline about the Supreme Court considering GOP-backed social media "censorship" laws, with mention of Florida's law. Background has the logos of Twitter and another large X icon. \\
X34 & debris; damaged; scattered; building; tornado; buildings; trees; rubble; area; house & Damaged house with a partially collapsed roof, surrounded by debris and scattered objects on a grassy area. Trees nearby, some bent or broken. Cloudy sky in the background.; Debris scattered on a lawn, a damaged house with a missing roof, fallen shingles, and an overturned piece of furniture. Damaged structures in the background. Text present.; Black SUV on partially collapsed driveway, exposed foundation with hanging pipes, tilted utility pole with cables. Text present. \\
X35 & images; eric; adams; displayed; corruption; bob; luther; martin; disney; king & Six framed photos on a blueprint background with names: Mike Lynch, Hannah Lynch (silhouette), Chris Morvillo, Neda Morvillo, Jonathan Bloomer, Judy Bloomer. Text below reads about five bodies recovered from a sunken yacht.; Portraits of six individuals marked as missing, including Mike Lynch, Jonathan Bloomer, Judy Bloomer, Chris Morvillo, Neda Morvillo, and an unidentified daughter. Text indicates yacht tragedy details.; Disney logo on blue background; text overlay mentions Bob Iger wins proxy fight as shareholders reelect full board. NBC News Now logo in the corner. \\
\bottomrule
\end{longtable}

\clearpage

\begin{figure}[H]
  \centering
  \includegraphics[width=0.8\textwidth]{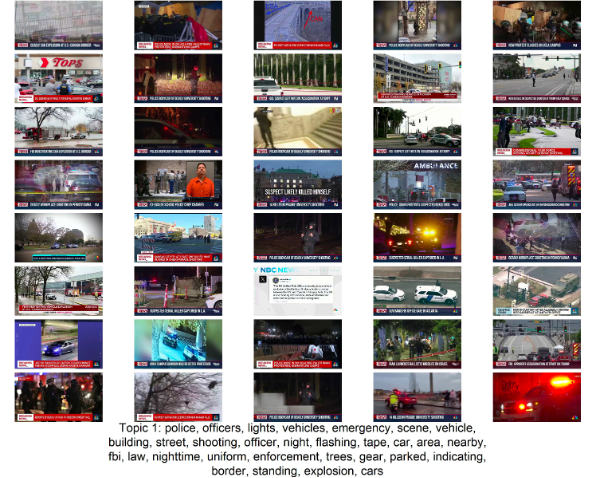} 
  \caption{Top Frames for Topic 1.}
  \label{fig:fig2}
\end{figure}

\begin{figure}{H}
  \centering
  \includegraphics[width=0.8\textwidth]{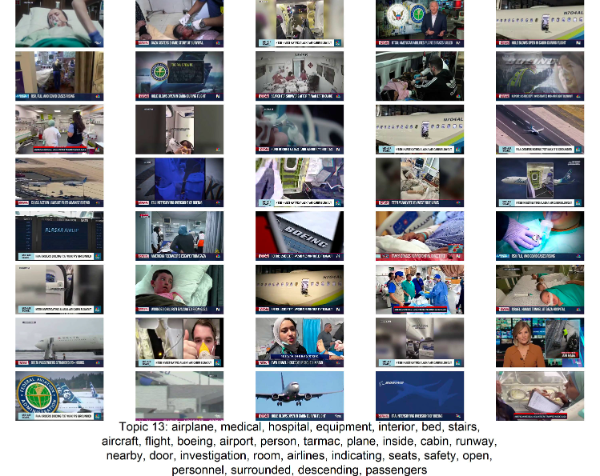} 
  \caption{Top Frames for Topic 13.}
  \label{fig:fig3}
\end{figure}

\begin{figure}{H}
  \centering
  \includegraphics[width=0.8\textwidth]{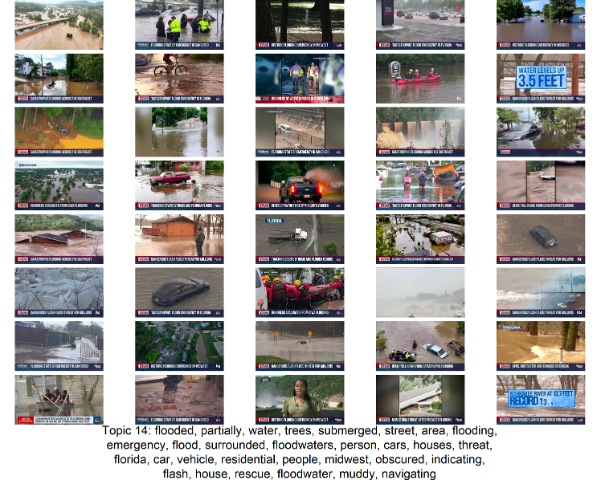} 
  \caption{Top Frames for Topic 14.}
  \label{fig:fig4}
\end{figure}

\begin{figure}{h}
  \centering
  \includegraphics[width=0.8\textwidth]{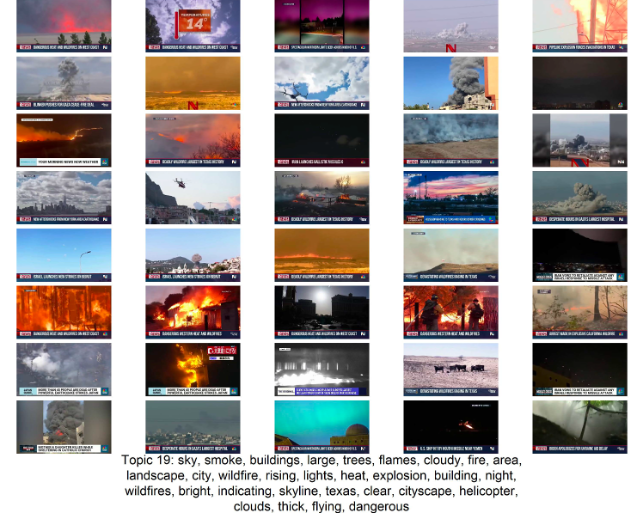} 
  \caption{Top Frames for Topic 19.}
  \label{fig:fig5}
\end{figure}

Representative frames from four randomly selected topics—Topics 1, 13, 14, and 19—are shown in Figures \ref{fig:fig2} to \ref{fig:fig5}, offering visual validation of the clustering results. These figures highlight the semantic coherence of the identified topics, which reflect both the thematic depth and framing strategies employed by NBC News.

Topic 1 centered on law enforcement scenes, dominated by visuals of police officers in riot gear, emergency vehicles, and urban crisis settings (see figure \ref{fig:fig2}). The associated captions, such as "Police officers in riot gear moving towards makeshift wooden structures" and "Scene of a car explosion near the U.S.-Canada border," align with these visuals to reinforce narratives of public safety and crisis management. Similarly, Topic 13 focused on medical emergencies, with frames depicting hospital interiors, medical personnel, and patient care (see figure \ref{fig:fig3}). Captions such as "A person lies in a hospital bed wearing an oxygen mask" and "Helicopter rescue operation over rocky terrain" highlight the dual emphasis on operational efficiency and the emotional stakes of healthcare crises.

Topic 14 illustrated the devastation of natural disasters, showcasing visuals of flooded streets, submerged buildings, and emergency responses (see figure \ref{fig:fig4}). Captions like "Flooded town with submerged streets and buildings" and "Rescue workers assisting residents through waterlogged streets" emphasize the infrastructural and human toll of these events. Meanwhile, Topic 19 captured the dramatic imagery of wildfires, featuring burning landscapes and smoky skies (see figure \ref{fig:fig5}). Frames are paired with captions such as "Wildfire with intense red and orange flames in the sky" and "Thermometer graphic with 14 degrees, cloudy sky," underscoring the urgency and environmental stakes of these crises.

Political themes were pervasive across the dataset, with topics centered on figures like Donald Trump and Joe Biden using visual motifs of leadership, authority, and patriotism. Frames of podiums, American flags, and campaign backdrops reinforced narratives of governance and power, shaping audience perceptions of these political figures. Similarly, clusters focused on crisis and emergency visuals highlighted dramatic and emotive imagery, such as burning forests and rescue operations, to frame events as both urgent and consequential. 

Beyond political and crisis-related themes, human-interest topics offered a more personal perspective, with clusters focused on family life, community gatherings, and everyday moments. These themes, which often employed warm and sentimental visuals, provided a narrative balance, demonstrating NBC News' ability to connect with audiences on a human level while diversifying its coverage. The dataset's thematic diversity, which also included topics like corporate branding and public protests, reflects the network’s commitment to covering a wide range of societal issues.

Validation of the clustering model underscored its robustness and interpretability. In the image intrusion test, coders correctly identified the "intruder" image in 75\% of cases (for both coders), demonstrating strong internal consistency within the clusters. Similarly, in the topic-matching test, coders successfully matched individual frames to their respective topics in 86\% (coder 1) and 87\% (coder 2) of cases, confirming alignment between visual and semantic themes. These findings illustrate the value of computational tools in systematically analyzing large-scale visual datasets, revealing nuanced framing strategies in contemporary news media. By employing these techniques, the study highlights how visuals are used not only to inform but also to influence audience perceptions and emotional responses.

\subsection{Limitations and Ethical Considerations}
While this study demonstrates the potential of integrating unsupervised clustering and large language models for visual data analysis, several limitations merit attention. First, the platform constraints of tools like FastDup restrict their accessibility. FastDup operates exclusively on Docker and Mac OS, limiting its utility for researchers working on other operating systems, such as Windows, without Docker environments. Additionally, while FastDup’s hash-based similarity detection efficiently reduces redundancy, it may inadvertently exclude frames with subtle yet meaningful differences, potentially impacting the dataset’s diversity and analytical depth.

The reliance on OpenAI’s large language models for image captioning introduces further challenges. The financial costs associated with these models are substantial, particularly when analyzing large-scale datasets containing thousands of frames or images. This constraint may hinder broader adoption in resource-limited settings. Furthermore, processing large datasets is time-intensive; in this study, captioning 6,928 frames required over 10 hours. Such demands can be prohibitive for workflows requiring rapid analysis or real-time processing. Beyond time and cost, the accuracy of generated captions may be affected by ambiguous visual contexts, necessitating iterative prompt refinement to align outputs with research goals.

Finally, the computational demands of full models, such as LDA or neural embeddings, highlight additional constraints. These models require significant storage for intermediate data, such as preprocessed frames and caption files, as well as the final outputs, particularly when analyzing dynamic video content. The computational time for tasks like frame extraction, deduplication, and topic modeling further limits scalability, especially for large, temporally rich datasets. These challenges underscore the need for streamlined workflows and scalable computational infrastructures to enable more efficient analysis while maintaining methodological rigor.

While the proposed method provides robust tools for visual data analysis, it also raises several ethical concerns that warrant careful consideration. The use of image captioning and clustering tools relies heavily on the data used to train these models, which may perpetuate biases inherent in the training datasets. For example, captions generated for politically charged or socially sensitive visuals could reflect underlying cultural or ideological biases, potentially distorting the interpretation of media narratives.

Privacy concerns are another critical issue, particularly when dealing with datasets containing images of identifiable individuals. The analysis of such data must adhere to ethical standards and legal frameworks, including obtaining consent when necessary and ensuring compliance with data protection regulations like GDPR. Additionally, automated analysis at scale may inadvertently decontextualize images, leading to misinterpretations or oversimplified conclusions that could misrepresent the original intent of the visuals.

There is also the potential for misuse of this technology in ways that reinforce harmful media practices. For example, clustering and captioning tools could be employed to selectively frame narratives or amplify sensational content, contributing to misinformation or polarization. Researchers and practitioners must exercise caution and maintain transparency about the limitations and potential biases of these methods, ensuring that their applications prioritize accuracy, fairness, and accountability.

\section{Discussion and Future Work}
Despite these limitations and ethical concerns, this study demonstrates the potential of integrating unsupervised clustering methods with large language models to address critical gaps in the computational analysis of visual data. Traditional methods of visual content analysis, often reliant on manual coding or static approaches, struggle to scale effectively to the vast volumes of imagery produced by contemporary media ecosystems. By proposing and applying a robust pipeline for visual clustering, this study offers a scalable and systematic approach to understanding the visual dimensions of media content.

The results illustrate how this method overcomes key limitations in current approaches to visual analysis with black-box methods. By leveraging advanced clustering techniques, the methodology facilitates the identification of coherent visual and semantic themes across large datasets. The integration of deduplication tools such as FastDup ensures computational efficiency without compromising dataset diversity, while image captioning via large language models enhances the semantic depth of the analysis. This dual-layered approach allows researchers to uncover latent thematic structures within visual data, bridging the gap between visual representation and semantic meaning, while allowing the researcher to choose their own clustering methodologies. 

The application of this method to NBC News videos demonstrates its ability to capture both the breadth and depth of visual narratives. The findings highlight how major media outlets employ visual framing strategies to influence audience perceptions, offering insights into the interplay between visual elements and their narrative context. From the political imagery of campaign rallies to the dramatic depictions of environmental crises, this method enables the systematic exploration of themes that are central to contemporary communication research.

To build upon this methodological contribution, several avenues for future research are proposed. First, the development of more accessible and cross-platform tools for image deduplication, such as those compatible with Windows, would enhance the inclusivity and utility of the pipeline. Improving the granularity of deduplication algorithms to retain subtle but meaningful visual variations could further enrich the diversity and analytical potential of visual datasets.

Future work could also explore the integration of alternative captioning models (such as Sales Force Bootstrapping Language-Image Pre-training; BLIP) to mitigate the cost and time-intensive nature of OpenAI's tools. Open-source models or hybrid approaches combining automated and human-in-the-loop methods may offer cost-effective and contextually accurate solutions. With the rapid development of LLM’s it is reasonable to expect the quality of image captioning offered at the time of this study by OpenAI’s tools to be matched by free open-source models in the future. Additionally, the use of multimodal models that combine textual, visual, and audio inputs may provide richer insights into dynamic video datasets. 

Scalability remains a critical challenge, particularly for large datasets with high temporal resolution. Research into optimized data storage and computational workflows, including parallel processing and cloud-based solutions, would be beneficial. Further advancements in the application of neural embeddings and transfer learning techniques could improve the interpretability and efficiency of clustering models, particularly when applied to cross-platform or multilingual datasets.

Finally, expanding the scope of applications to include longitudinal studies of visual framing and comparative analyses across media outlets or geographic regions could provide valuable insights into evolving patterns of media representation. By addressing these areas, future research can continue to refine and expand the use of computational methods for visual analysis, contributing to a deeper understanding of the intersection between media, technology, and public discourse.

\textbf{Data availability statement:} The data that support the findings of this study are openly available in Open Science Framework at [redacted for review], reference number [redacted for review]. The python package for data analysis will be available at pypi [package name redacted for review], post publication. 


\clearpage

\section{Supplementary Material} \label{Supplement}
\beginsupplement
\subsection{List of Packages} \label{SupA}

\subsubsection{Python Packages}
\begin{itemize}
    \item \href{https://pillow.readthedocs.io/en/stable/}{PIL (Pillow)}
    \item \href{https://docs.python.org/3/library/base64.html}{base64}
    \item \href{https://docs.python.org/3/library/concurrent.html}{concurrent}
    \item \href{https://docs.python.org/3/library/csv.html}{csv}
    \item \href{https://docs.opencv.org/}{cv2 (OpenCV)}
    \item \href{https://docs.python.org/3/library/datetime.html}{datetime}
    \item \href{https://visual-layer.github.io/fastdup/}{fastdup}
    \item \href{https://docs.python.org/3/library/glob.html}{glob}
    \item \href{https://gradio.app/docs/}{gradio}
    \item \href{https://docs.python.org/3/library/io.html}{io}
    \item \href{https://docs.python.org/3/library/logging.html}{logging}
    \item \href{https://numpy.org/doc/stable/}{numpy}
    \item \href{https://platform.openai.com/docs}{openai}
    \item \href{https://docs.python.org/3/library/os.html}{os}
    \item \href{https://pandas.pydata.org/docs/}{pandas}
    \item \href{https://docs.python.org/3/library/random.html}{random}
    \item \href{https://docs.python.org/3/library/re.html}{re}
    \item \href{https://docs.python-requests.org/en/latest/}{requests}
    \item \href{https://docs.python.org/3/library/subprocess.html}{subprocess}
    \item \href{https://docs.python.org/3/library/sys.html}{sys}
    \item \href{https://docs.python.org/3/library/threading.html}{threading}
    \item \href{https://docs.python.org/3/library/time.html}{time}
    \item \href{https://docs.python.org/3/library/tkinter.html}{tkinter}
    \item \href{https://github.com/yt-dlp/yt-dlp}{yt\_dlp}
\end{itemize}

\subsubsection{R Packages}
\begin{itemize}
    \item cld2
    \item corpustools
    \item daiR
    \item devtools
    \item doParallel
    \item dplyr
    \item foreign
    \item ggplot2
    \item igraph
    \item irr
    \item ldatuning
    \item lsa
    \item lubridate
    \item magick
    \item parallel
    \item quanteda
    \item ragree
    \item readr
    \item scales
    \item stopwords
    \item stringi
    \item stringr
    \item tidyr
    \item tidytext
    \item tidyverse
    \item topicmodels
    \item xlsx
\end{itemize}

\clearpage
\subsection{LDA Results} \label{SubB}

\begin{figure}[H]
  \centering
  \includegraphics[width=0.7\textwidth]{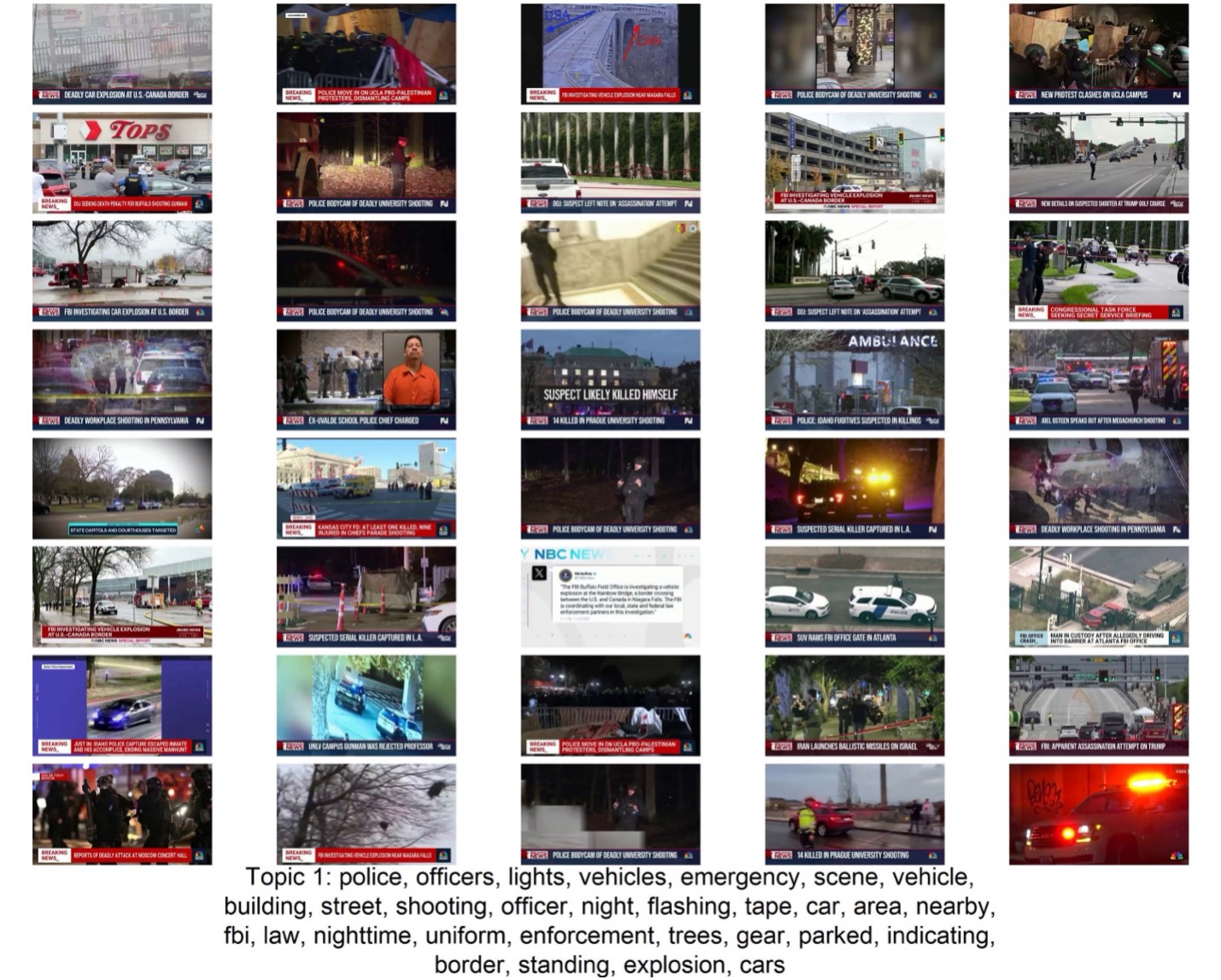} 
  \caption{Topic 1 Clustering Results.}
  \label{fig:figb1}
\end{figure}

\begin{figure}[H]
  \centering
  \includegraphics[width=0.7\textwidth]{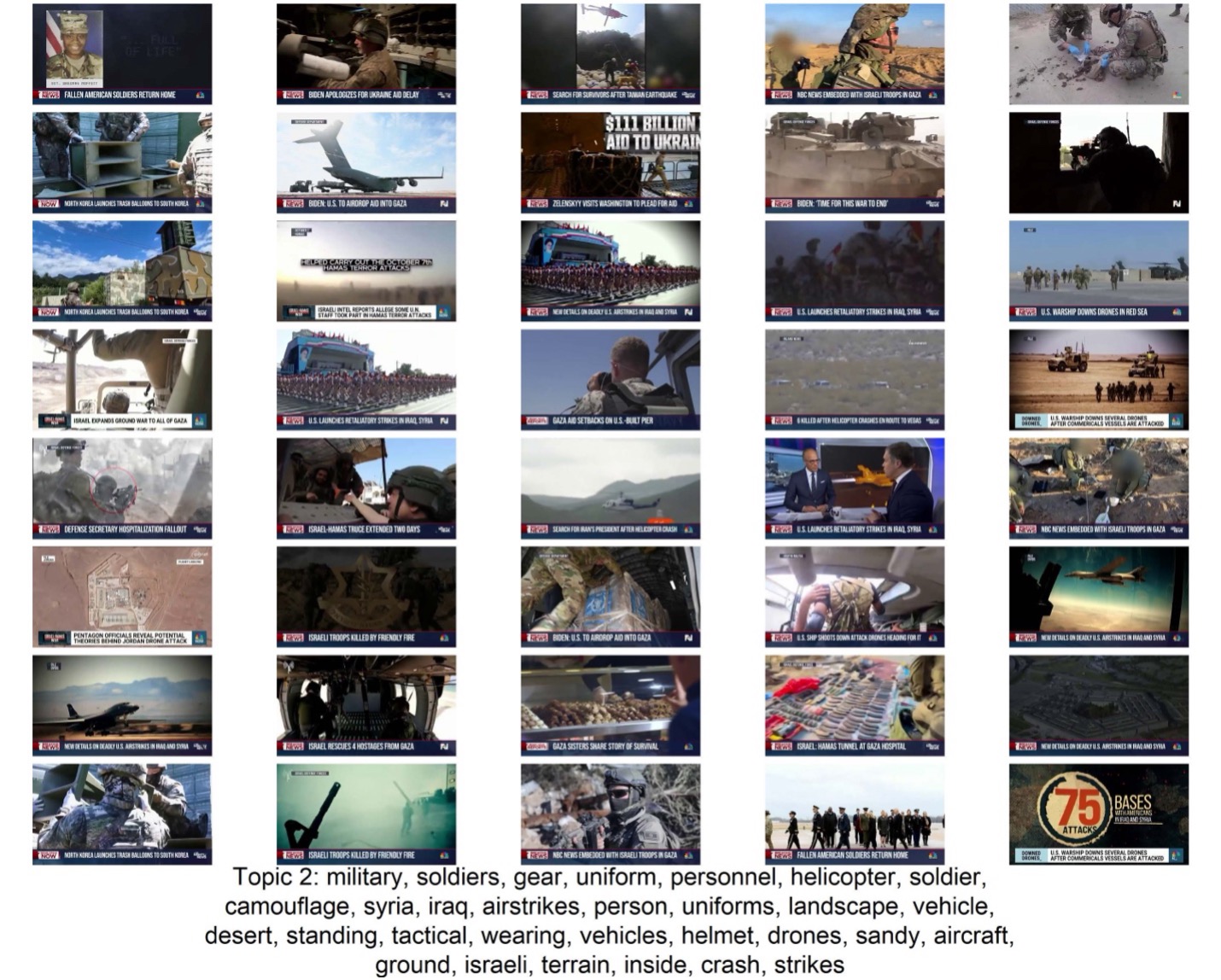} 
  \caption{Topic 2 Clustering Results.}
  \label{fig:figb2}
\end{figure}

\clearpage
\begin{figure}
  \centering
  \includegraphics[width=0.7\textwidth]{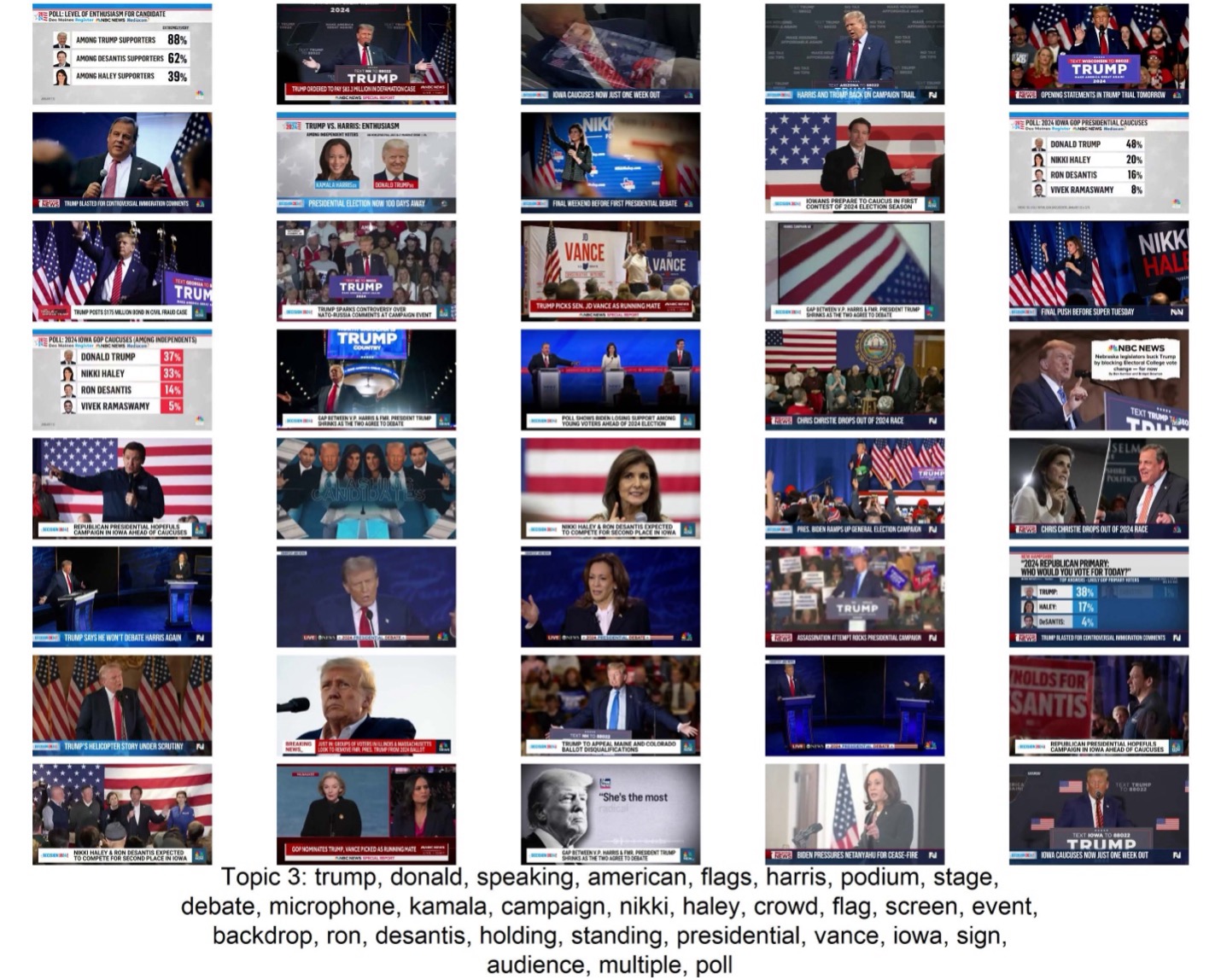} 
  \caption{Topic 3 Clustering Results.}
  \label{fig:figb3}
\end{figure}

\begin{figure}
  \centering
  \includegraphics[width=0.7\textwidth]{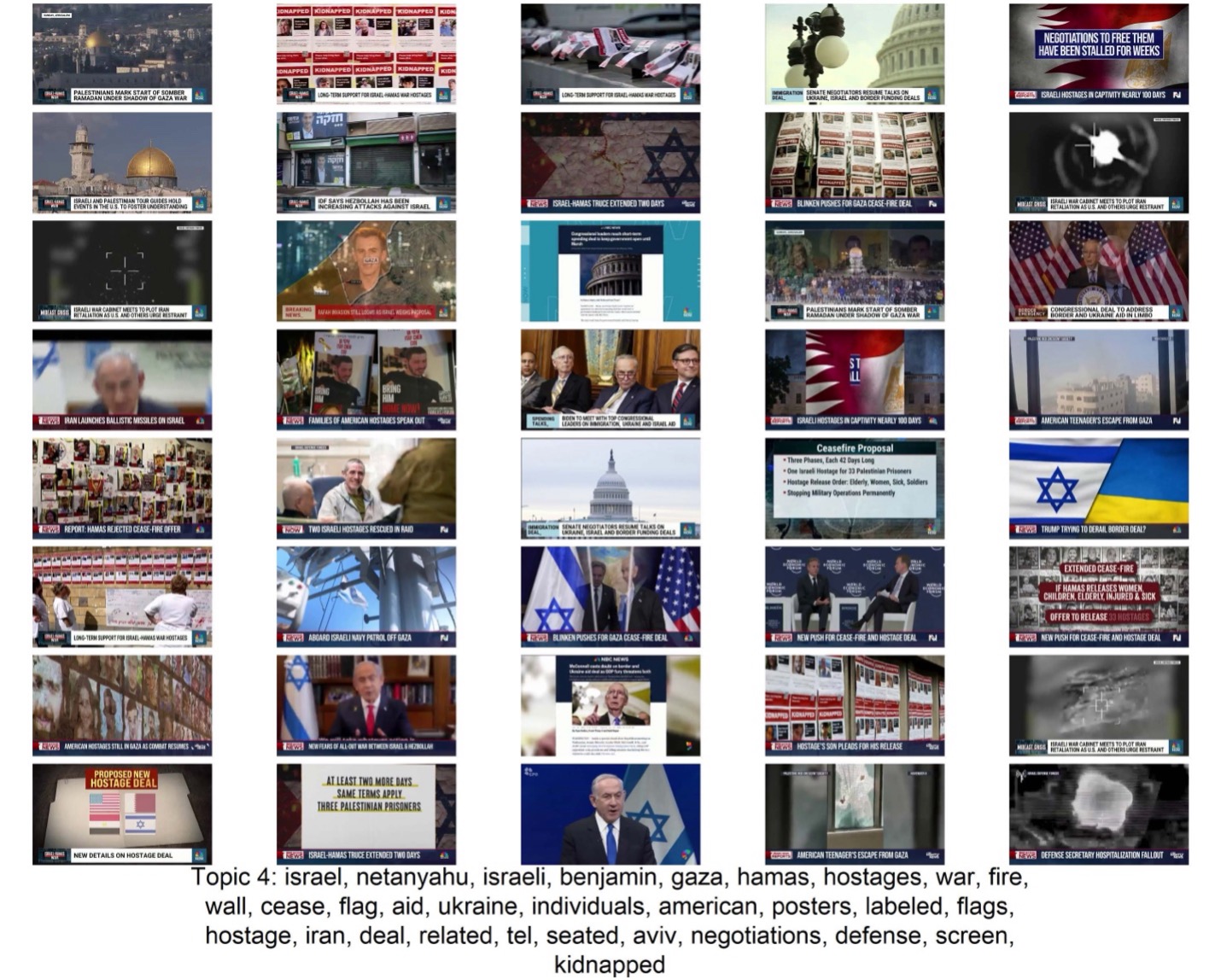} 
  \caption{Topic 4 Clustering Results.}
  \label{fig:figb4}
\end{figure}

\clearpage
\begin{figure}
  \centering
  \includegraphics[width=0.7\textwidth]{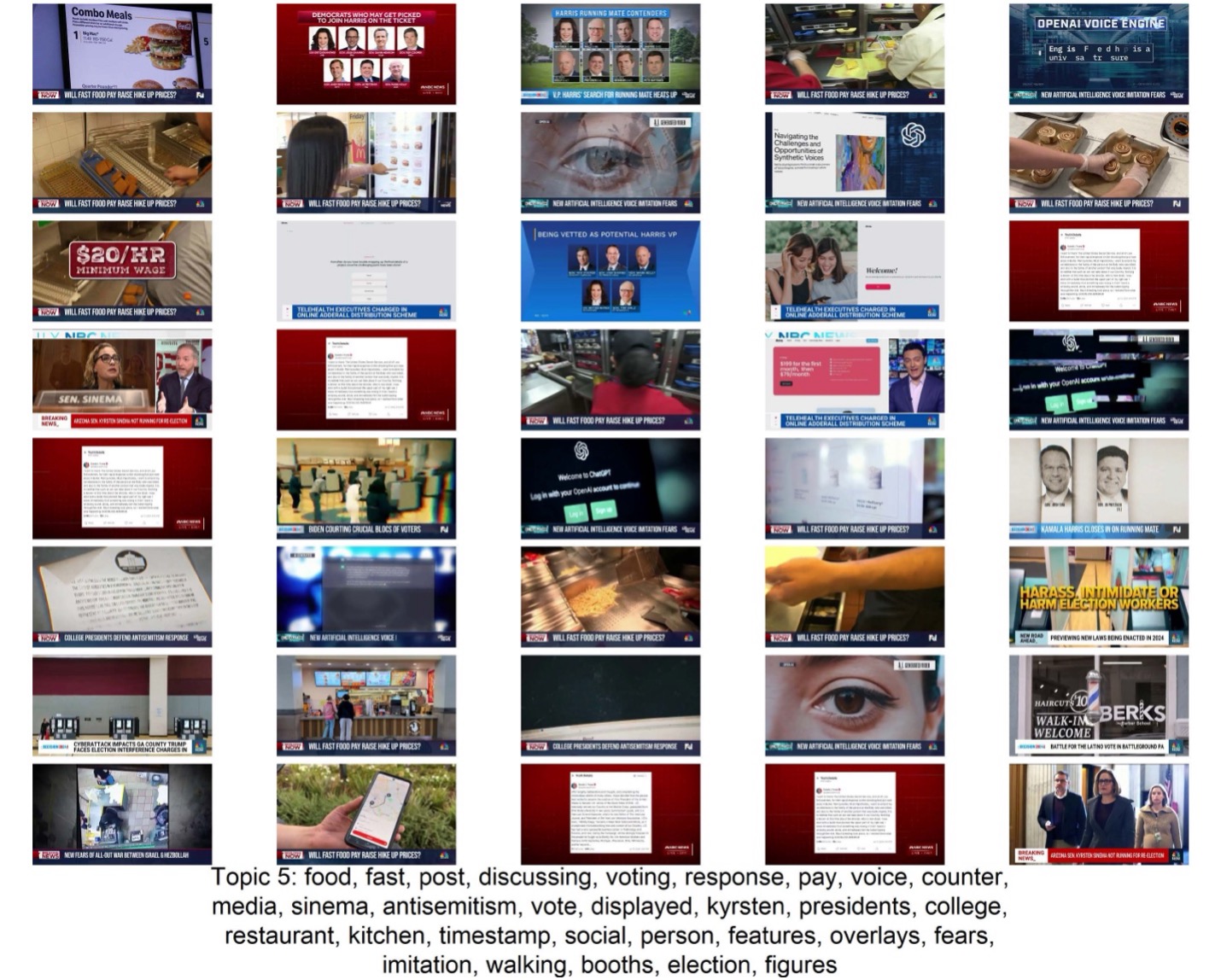} 
  \caption{Topic 5 Clustering Results.}
  \label{fig:figb5}
\end{figure}

\begin{figure}
  \centering
  \includegraphics[width=0.7\textwidth]{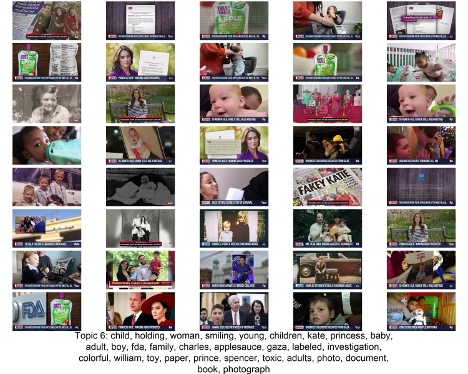} 
  \caption{Topic 6 Clustering Results.}
  \label{fig:figb6}
\end{figure}

\clearpage
\begin{figure}
  \centering
  \includegraphics[width=0.7\textwidth]{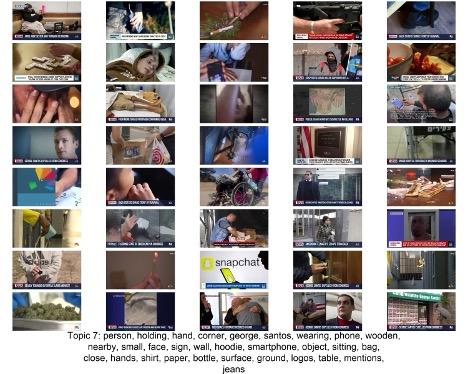} 
  \caption{Topic 7 Clustering Results.}
  \label{fig:figb7}
\end{figure}

\begin{figure}
  \centering
  \includegraphics[width=0.7\textwidth]{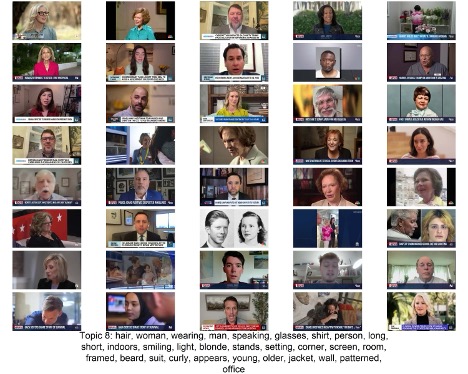} 
  \caption{Topic 8 Clustering Results.}
  \label{fig:figb8}
\end{figure}

\clearpage
\begin{figure}
  \centering
  \includegraphics[width=0.7\textwidth]{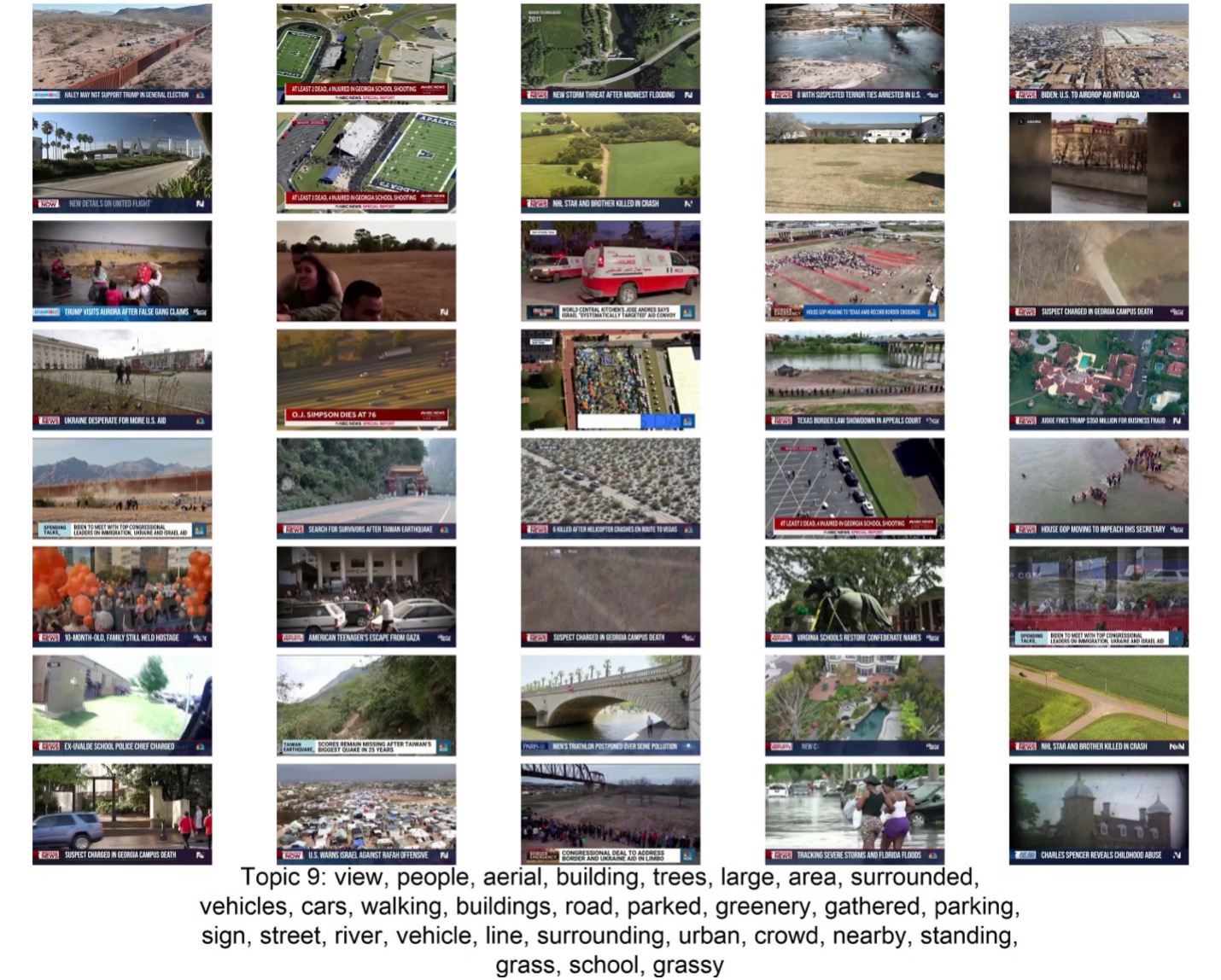} 
  \caption{Topic 9 Clustering Results.}
  \label{fig:figb9}
\end{figure}

\begin{figure}
  \centering
  \includegraphics[width=0.7\textwidth]{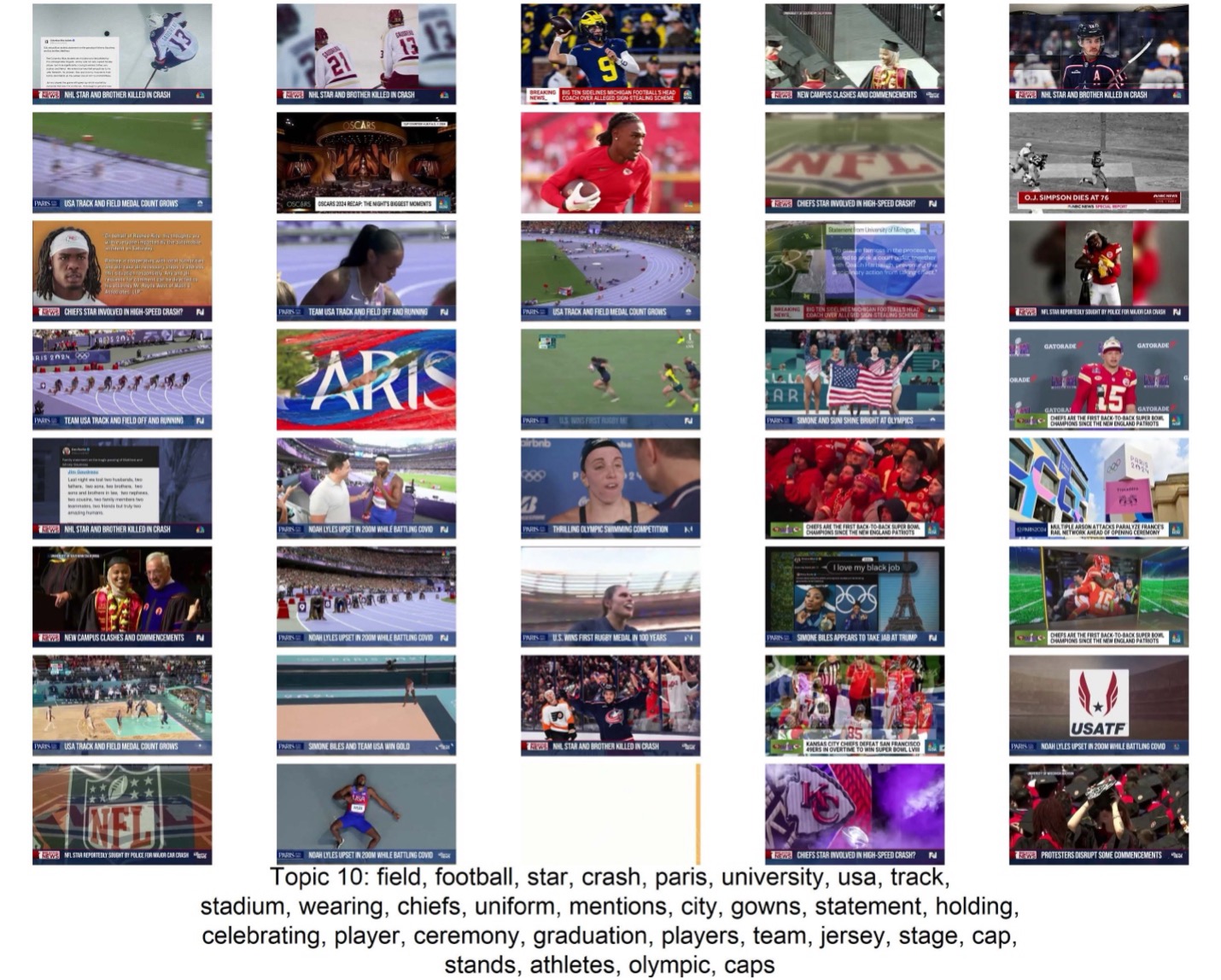} 
  \caption{Topic 10 Clustering Results.}
  \label{fig:figb10}
\end{figure}

\clearpage
\begin{figure}
  \centering
  \includegraphics[width=0.7\textwidth]{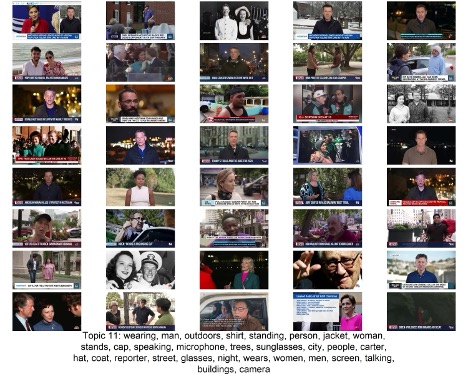} 
  \caption{Topic 11 Clustering Results.}
  \label{fig:figb11}
\end{figure}

\begin{figure}
  \centering
  \includegraphics[width=0.7\textwidth]{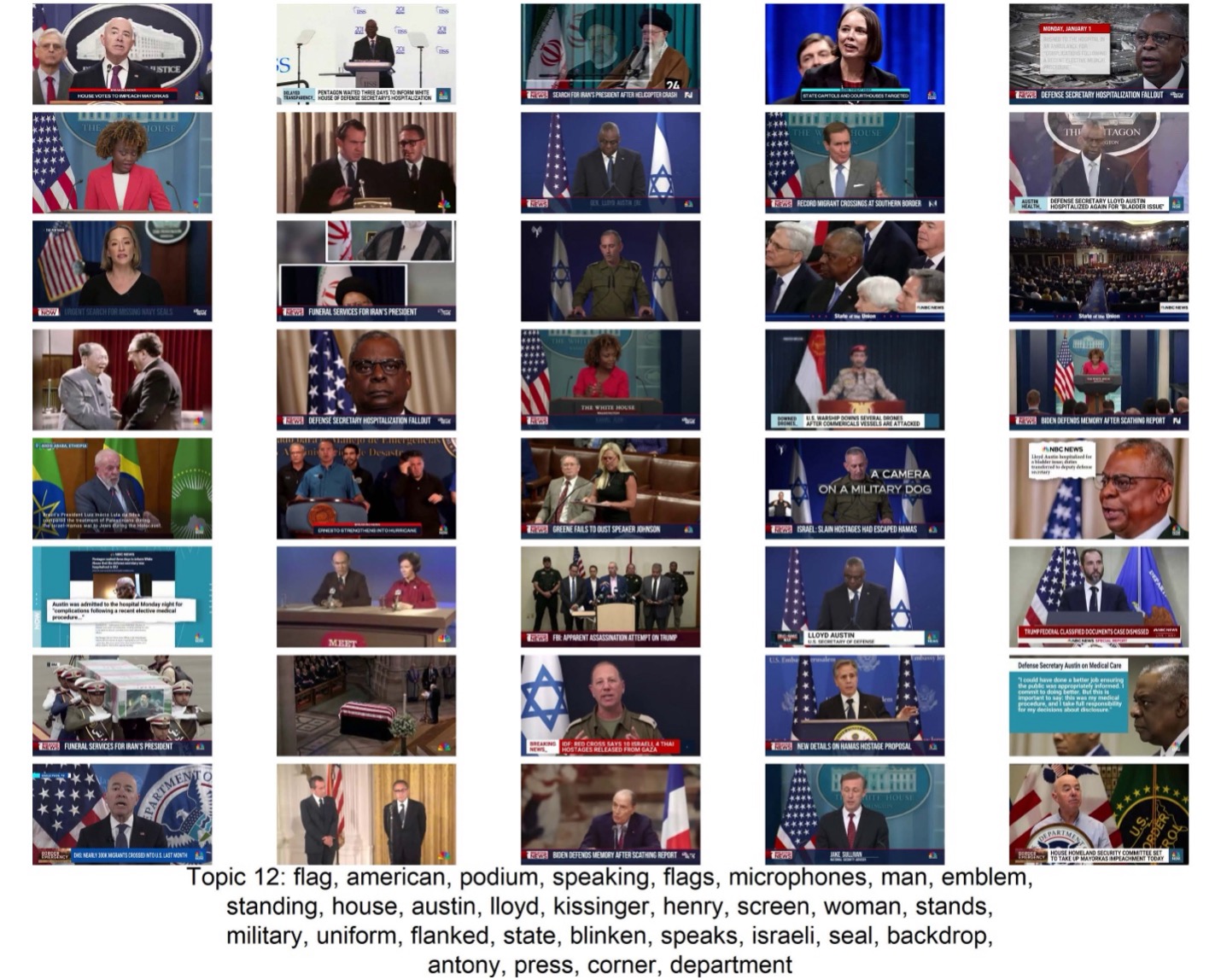} 
  \caption{Topic 12 Clustering Results.}
  \label{fig:figb12}
\end{figure}

\clearpage
\begin{figure}
  \centering
  \includegraphics[width=0.7\textwidth]{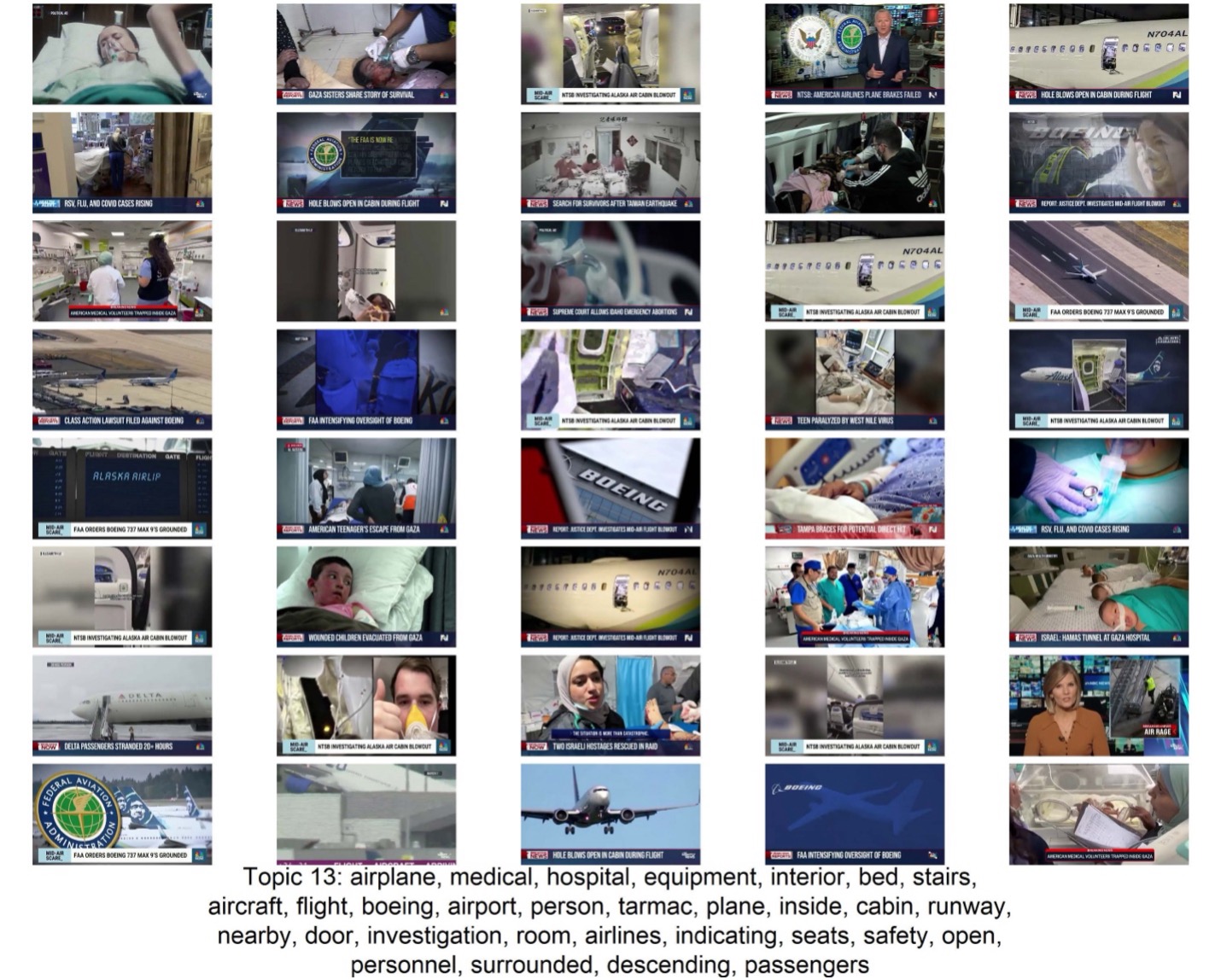} 
  \caption{Topic 13 Clustering Results.}
  \label{fig:figb13}
\end{figure}

\begin{figure}
  \centering
  \includegraphics[width=0.7\textwidth]{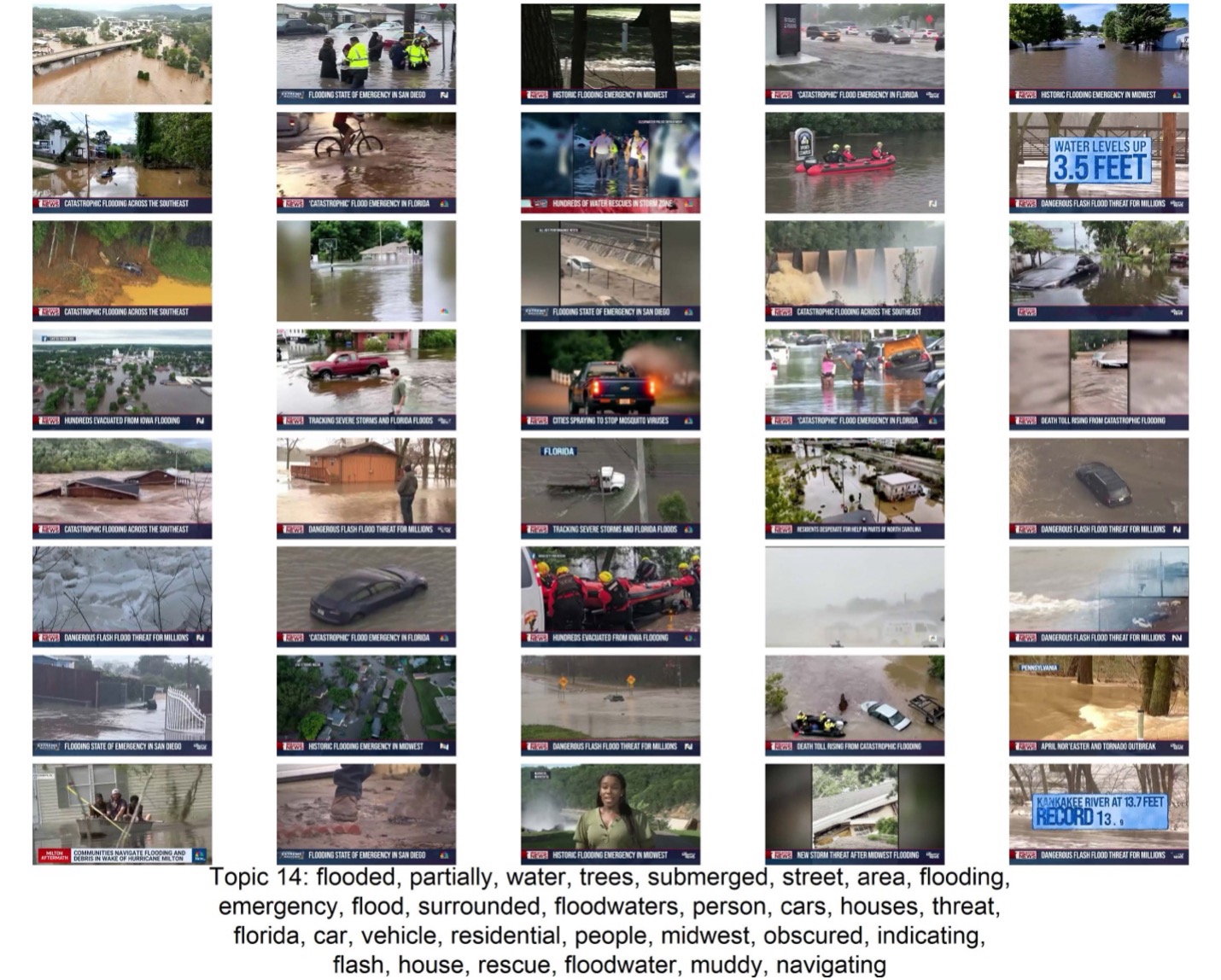} 
  \caption{Topic 14 Clustering Results.}
  \label{fig:figb14}
\end{figure}

\clearpage
\begin{figure}
  \centering
  \includegraphics[width=0.7\textwidth]{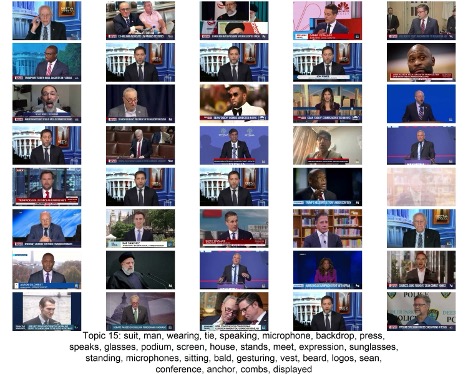} 
  \caption{Topic 15 Clustering Results.}
  \label{fig:figb15}
\end{figure}

\begin{figure}
  \centering
  \includegraphics[width=0.7\textwidth]{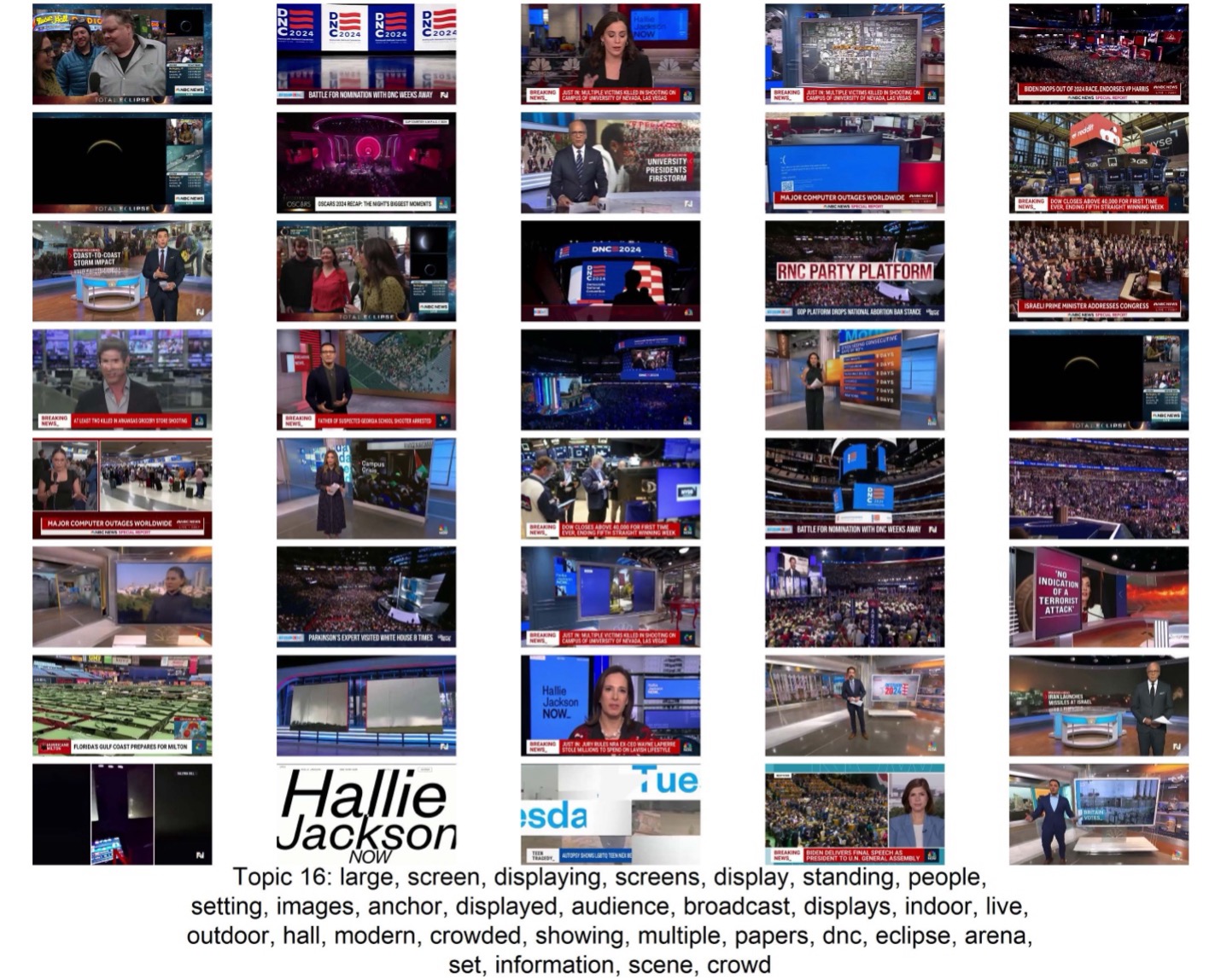} 
  \caption{Topic 16 Clustering Results.}
  \label{fig:figb16}
\end{figure}

\clearpage
\begin{figure}
  \centering
  \includegraphics[width=0.7\textwidth]{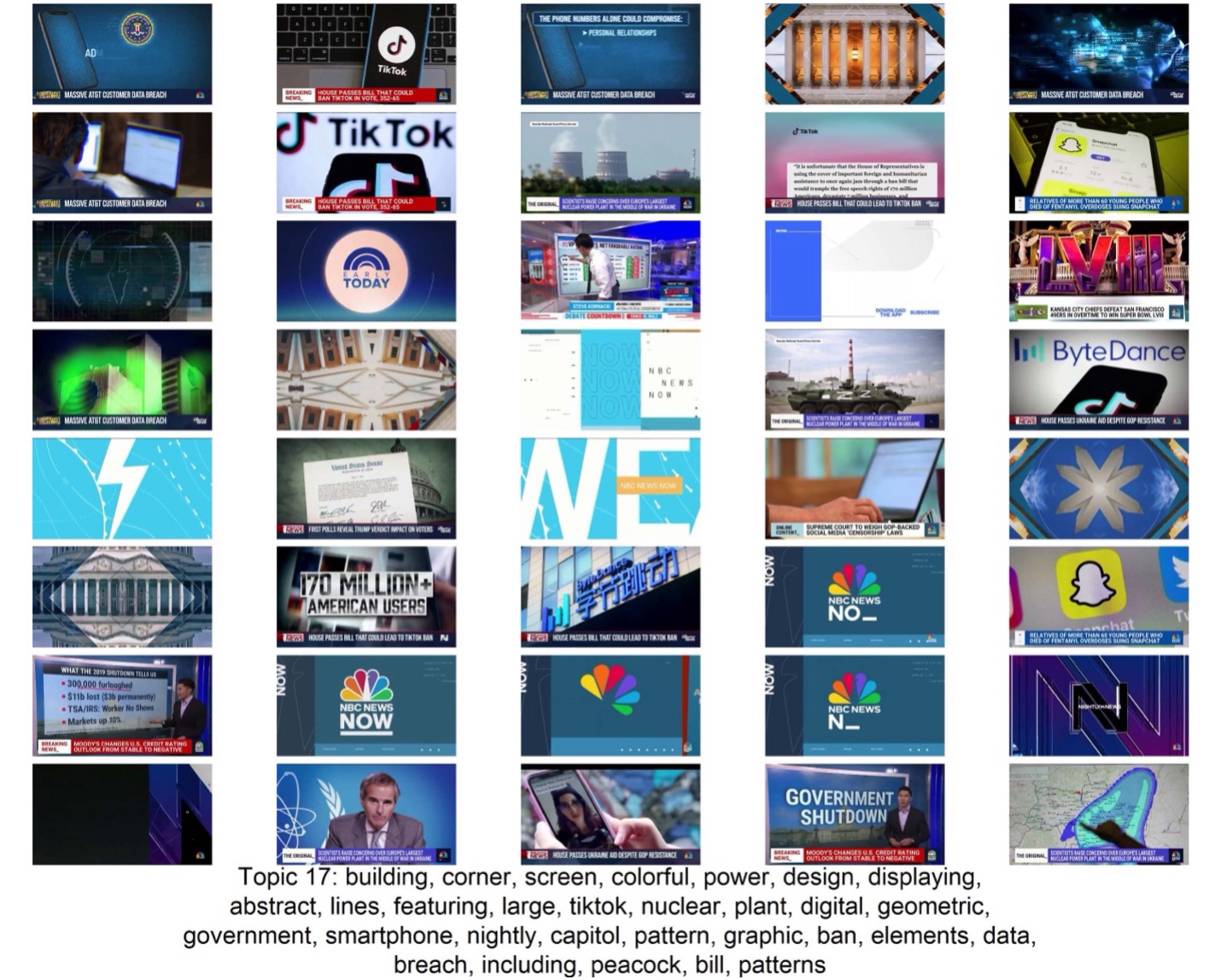} 
  \caption{Topic 17 Clustering Results.}
  \label{fig:figb17}
\end{figure}

\begin{figure}
  \centering
  \includegraphics[width=0.7\textwidth]{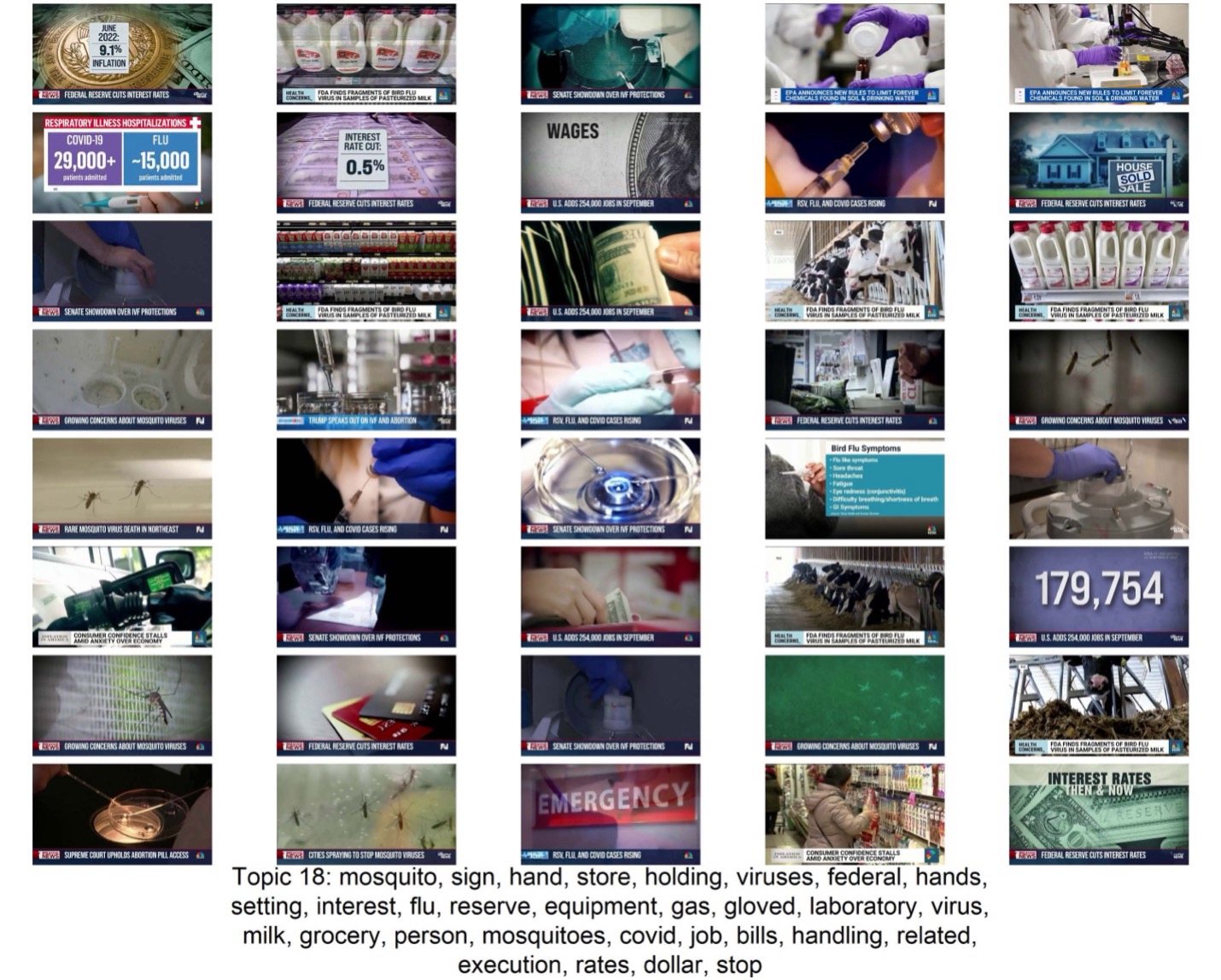} 
  \caption{Topic 18 Clustering Results.}
  \label{fig:figb18}
\end{figure}

\clearpage
\begin{figure}
  \centering
  \includegraphics[width=0.7\textwidth]{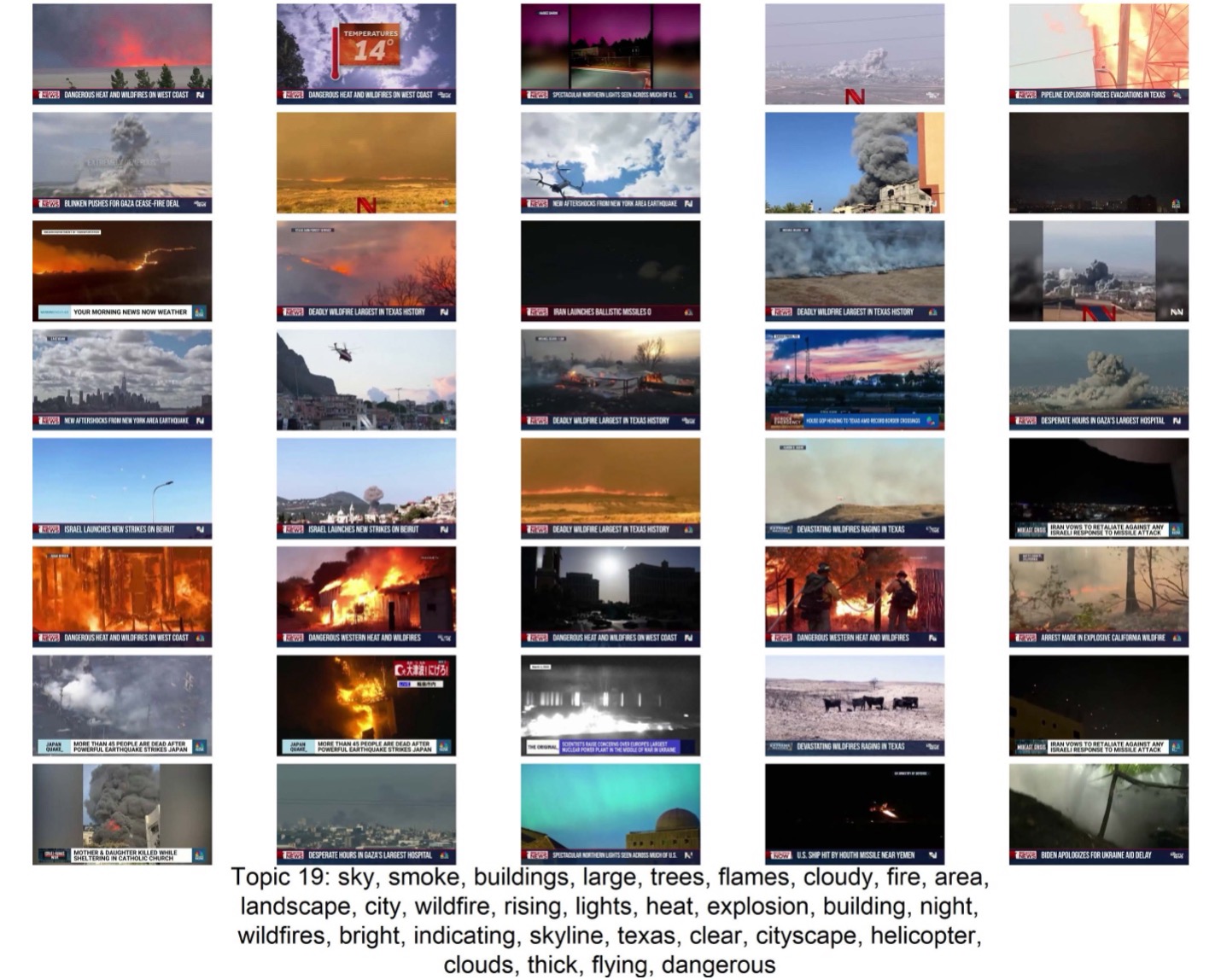} 
  \caption{Topic 19 Clustering Results.}
  \label{fig:figb19}
\end{figure}

\begin{figure}
  \centering
  \includegraphics[width=0.7\textwidth]{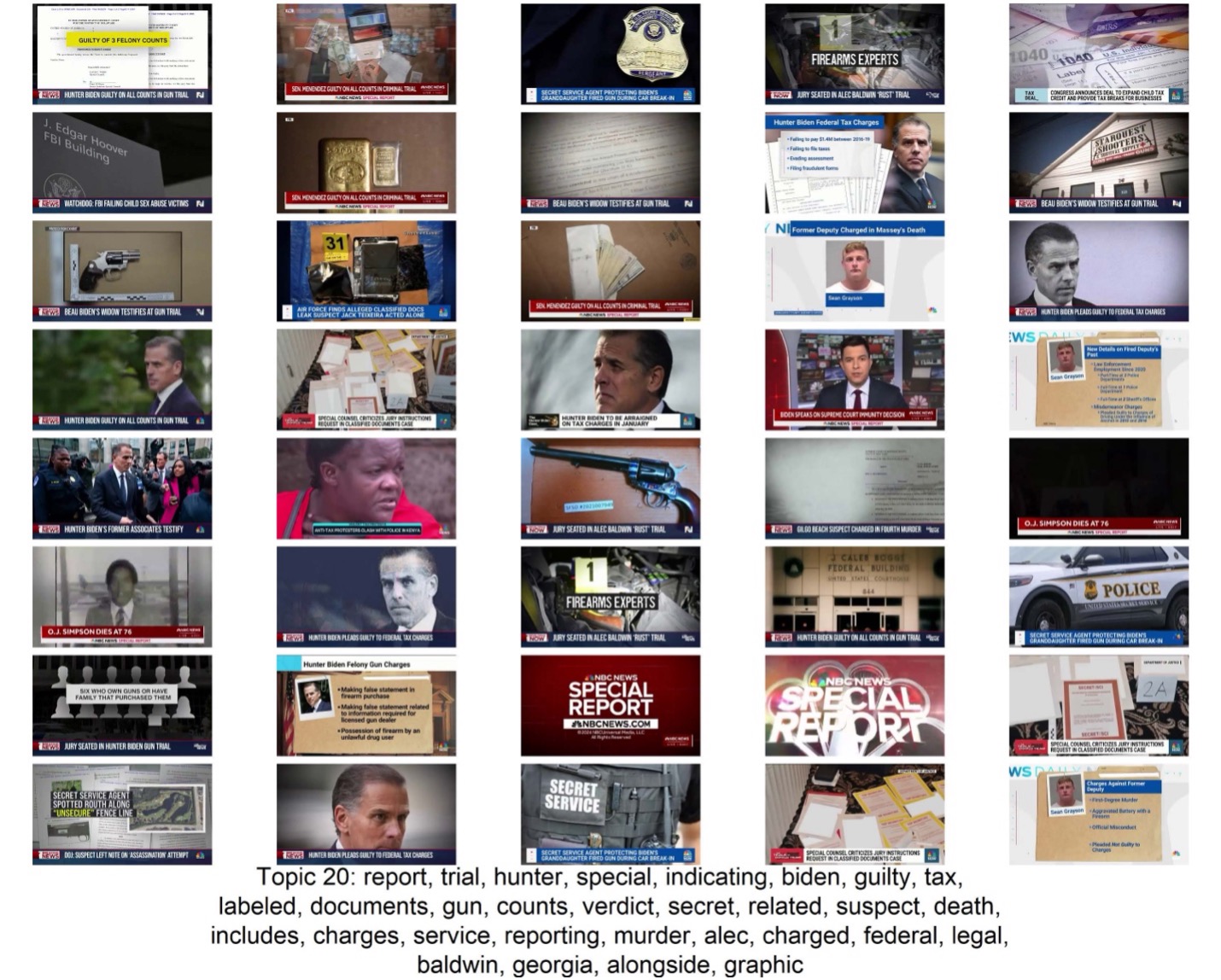} 
  \caption{Topic 20 Clustering Results.}
  \label{fig:figb20}
\end{figure}

\clearpage
\begin{figure}
  \centering
  \includegraphics[width=0.7\textwidth]{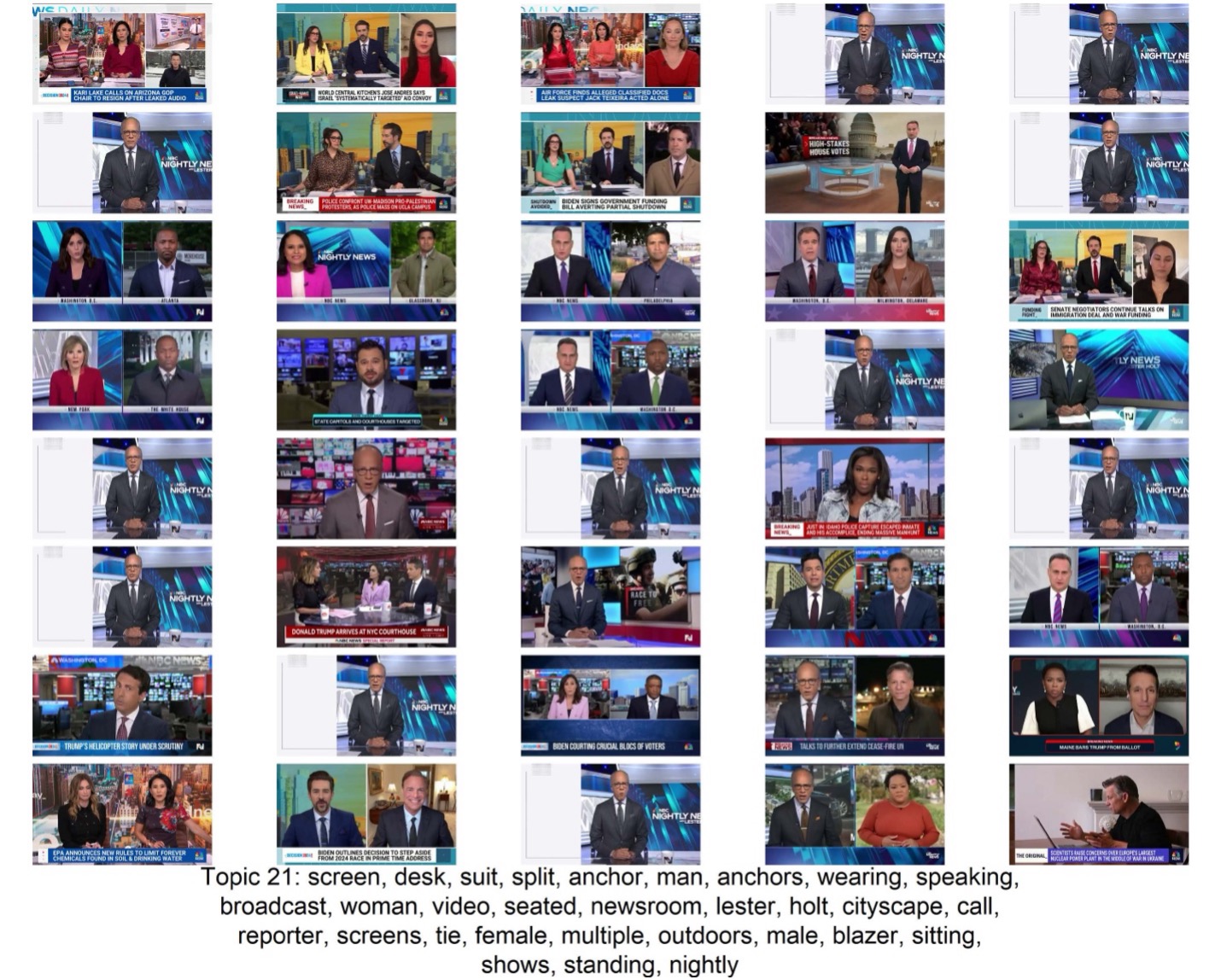} 
  \caption{Topic 21 Clustering Results.}
  \label{fig:figb21}
\end{figure}

\begin{figure}
  \centering
  \includegraphics[width=0.7\textwidth]{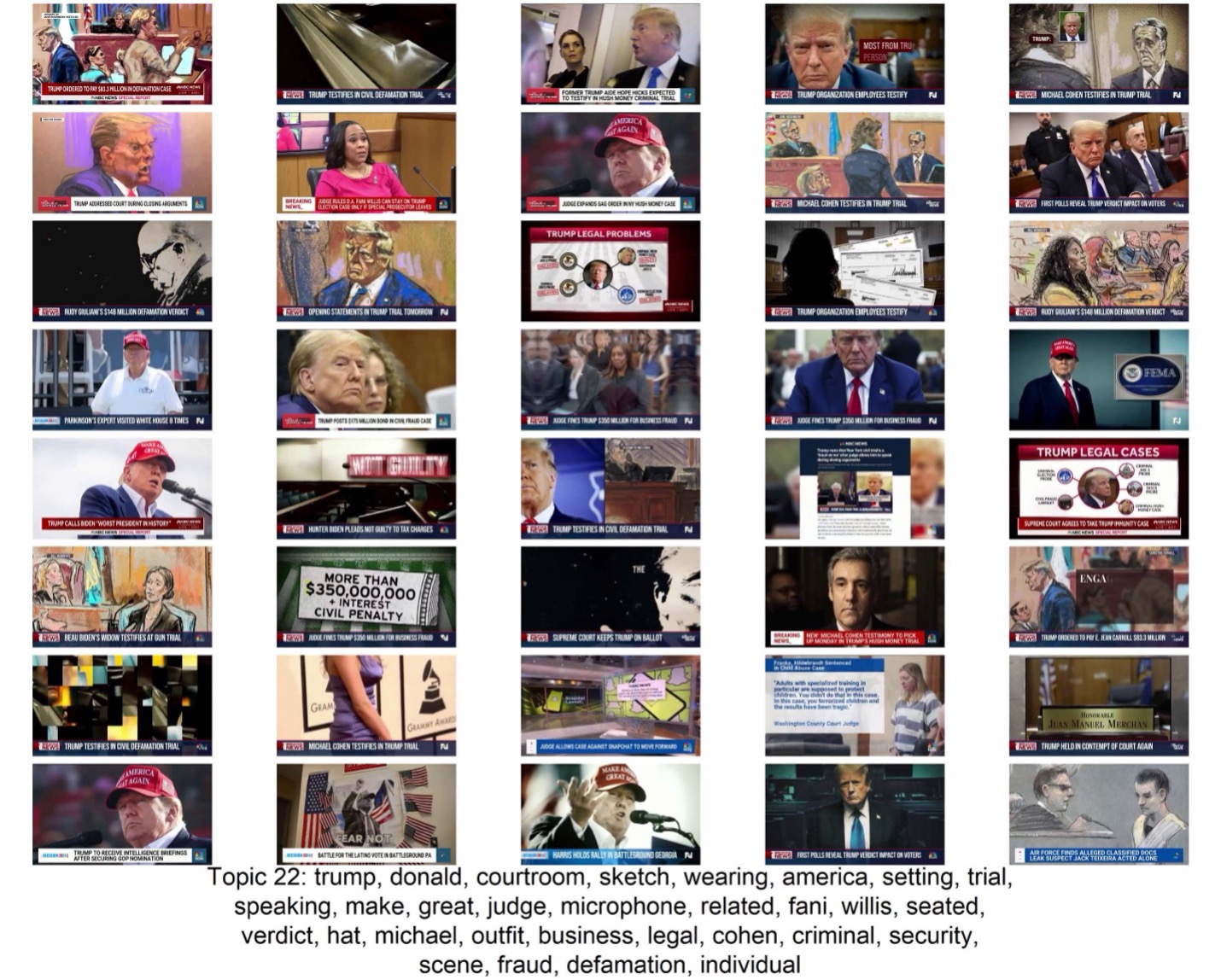} 
  \caption{Topic 22 Clustering Results.}
  \label{fig:figb22}
\end{figure}

\clearpage
\begin{figure}
  \centering
  \includegraphics[width=0.7\textwidth]{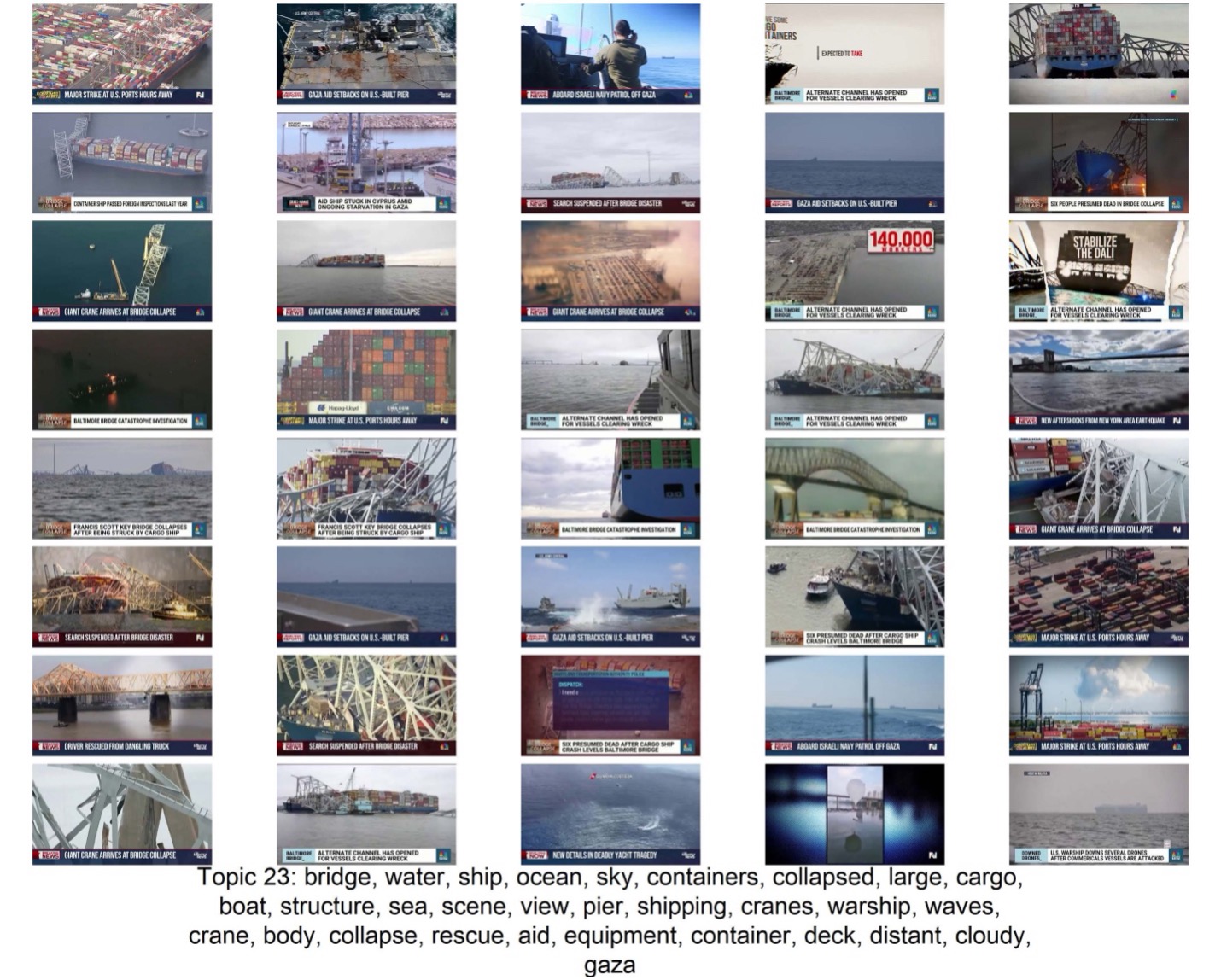} 
  \caption{Topic 23 Clustering Results.}
  \label{fig:figb23}
\end{figure}

\begin{figure}
  \centering
  \includegraphics[width=0.7\textwidth]{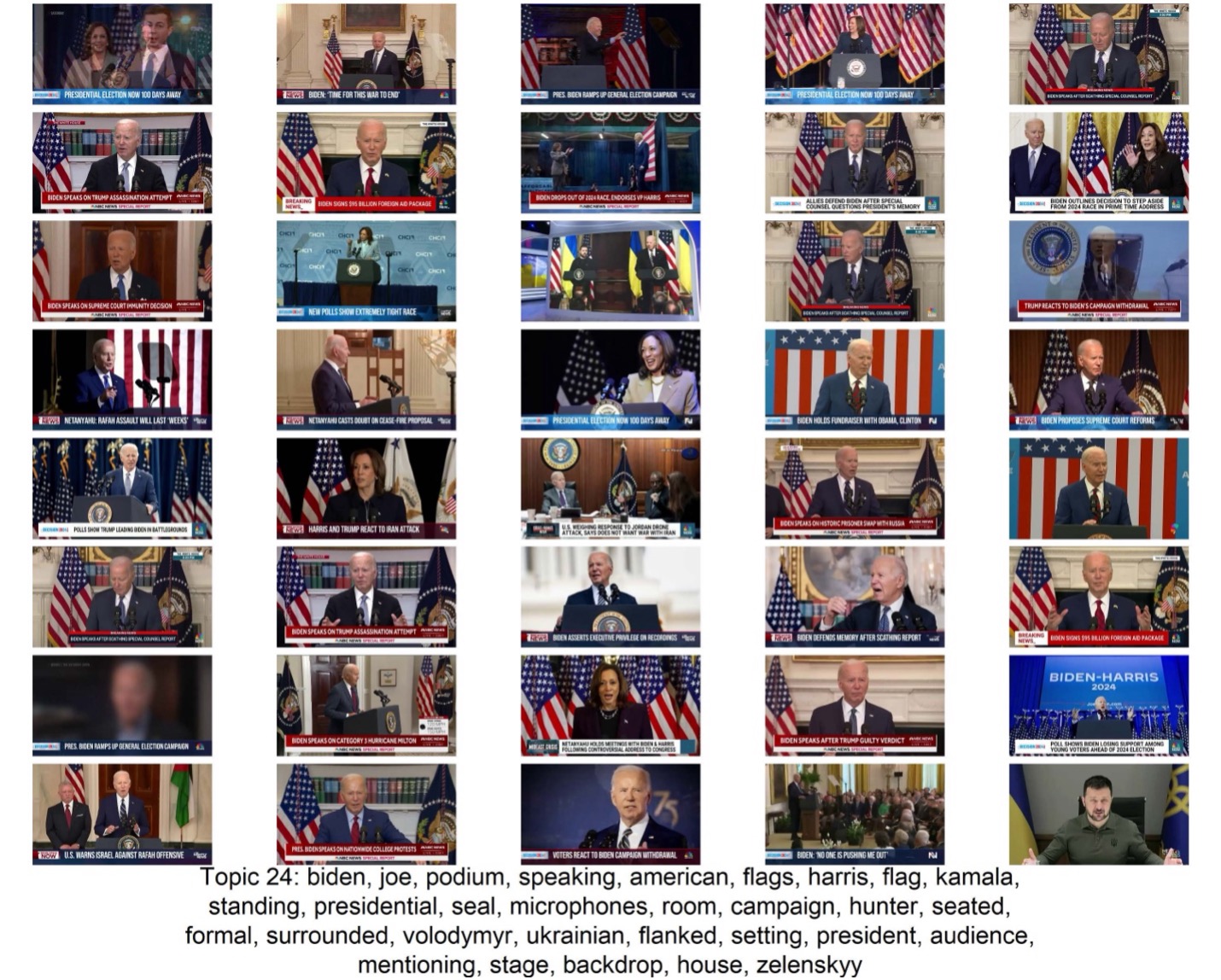} 
  \caption{Topic 24 Clustering Results.}
  \label{fig:figb24}
\end{figure}

\clearpage
\begin{figure}
  \centering
  \includegraphics[width=0.7\textwidth]{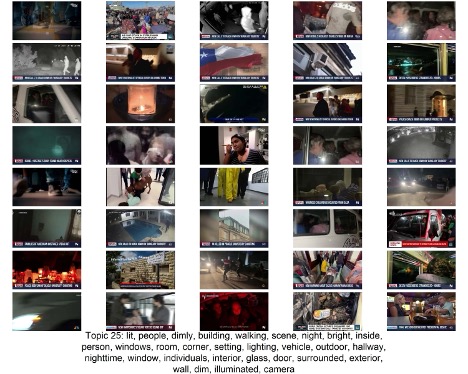} 
  \caption{Topic 25 Clustering Results.}
  \label{fig:figb25}
\end{figure}

\begin{figure}
  \centering
  \includegraphics[width=0.7\textwidth]{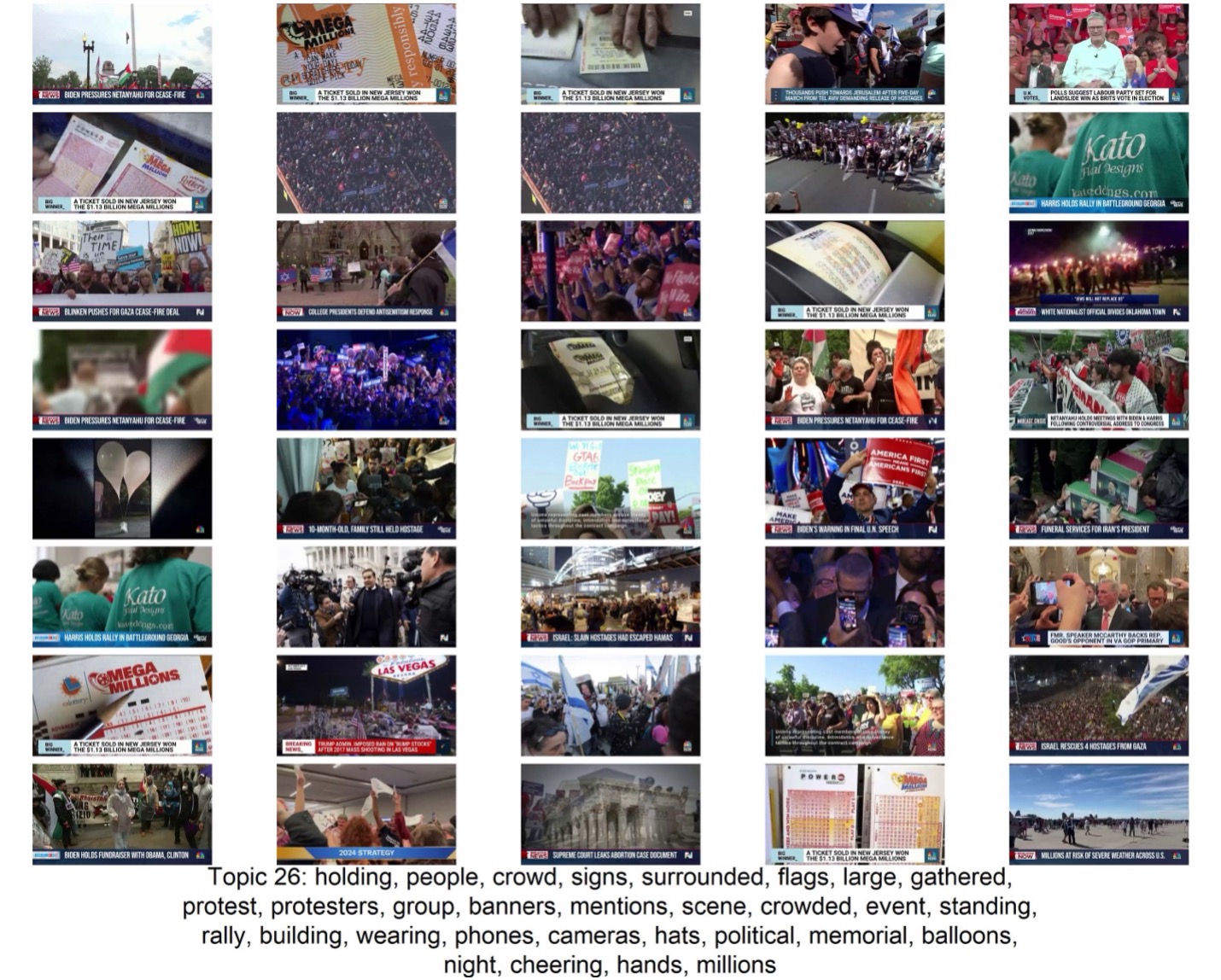} 
  \caption{Topic 26 Clustering Results.}
  \label{fig:figb26}
\end{figure}

\clearpage
\begin{figure}
  \centering
  \includegraphics[width=0.7\textwidth]{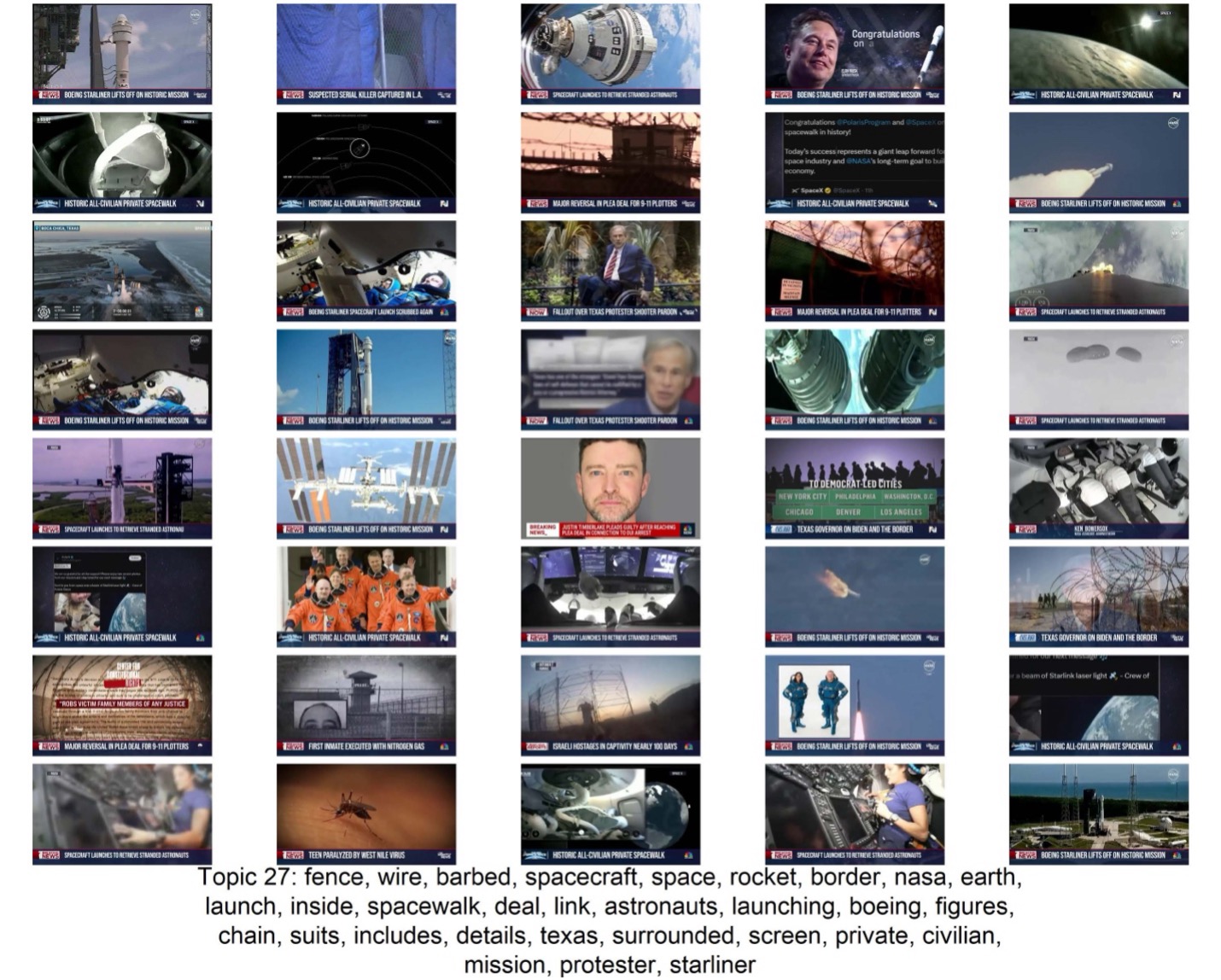} 
  \caption{Topic 27 Clustering Results.}
  \label{fig:figb27}
\end{figure}

\begin{figure}
  \centering
  \includegraphics[width=0.7\textwidth]{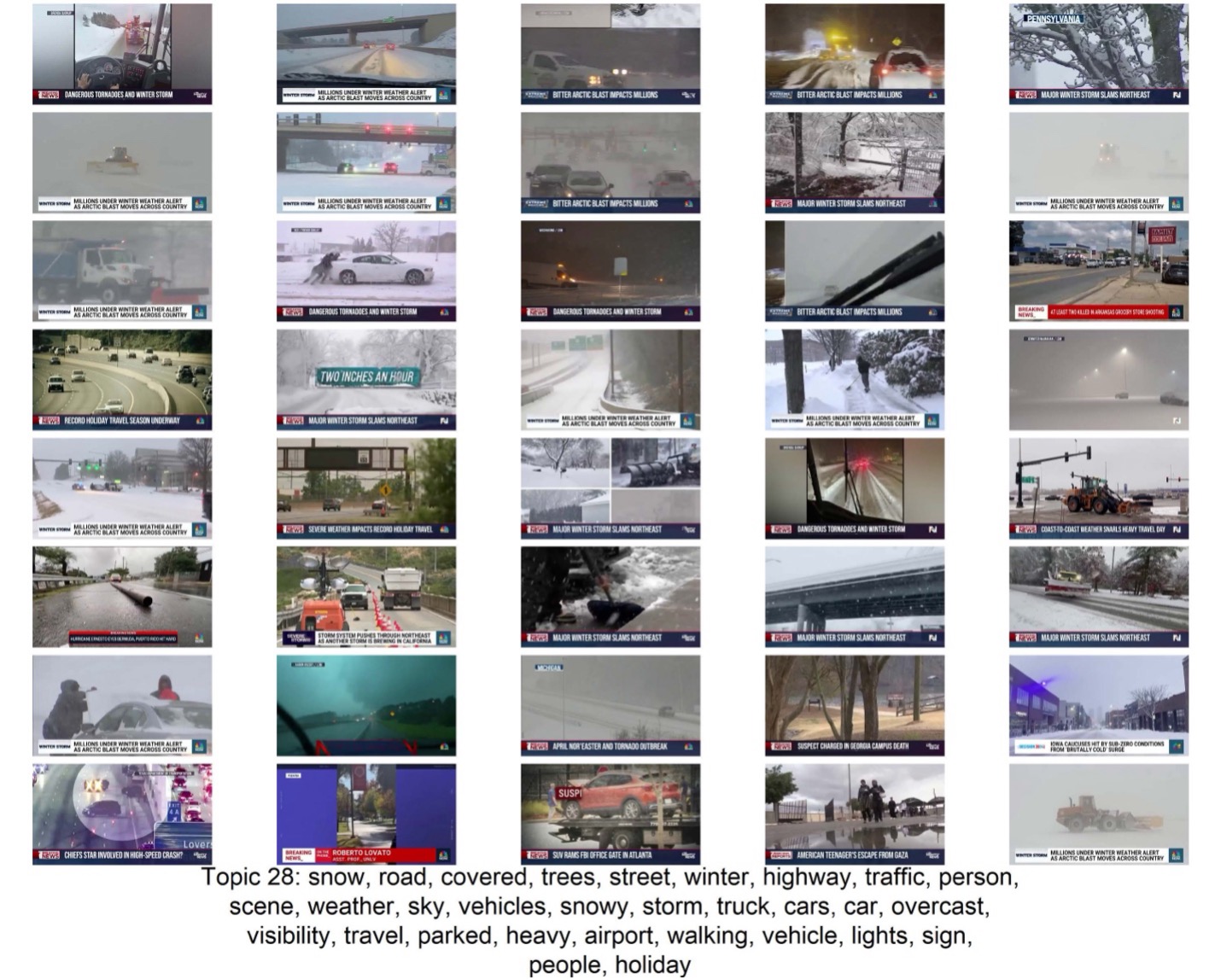} 
  \caption{Topic 28 Clustering Results.}
  \label{fig:figb28}
\end{figure}

\clearpage
\begin{figure}
  \centering
  \includegraphics[width=0.7\textwidth]{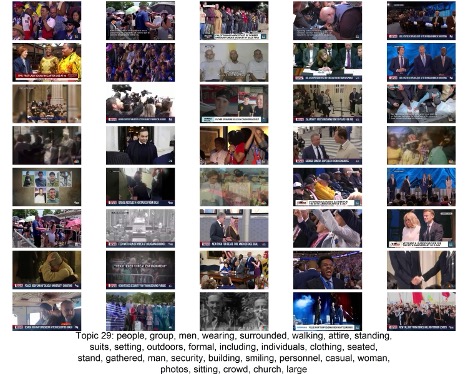} 
  \caption{Topic 29 Clustering Results.}
  \label{fig:figb29}
\end{figure}

\begin{figure}
  \centering
  \includegraphics[width=0.7\textwidth]{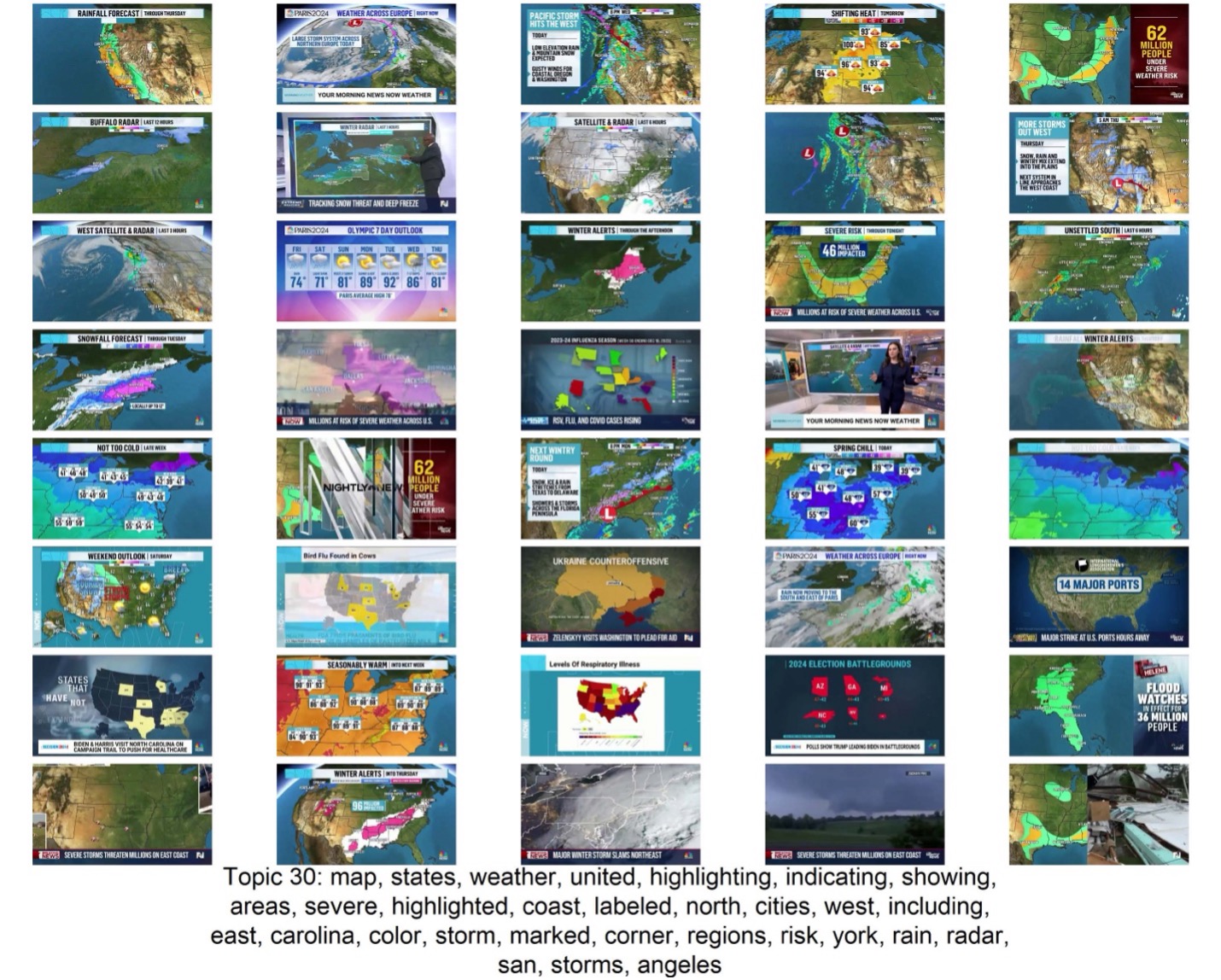} 
  \caption{Topic 30 Clustering Results.}
  \label{fig:figb30}
\end{figure}

\clearpage
\begin{figure}
  \centering
  \includegraphics[width=0.7\textwidth]{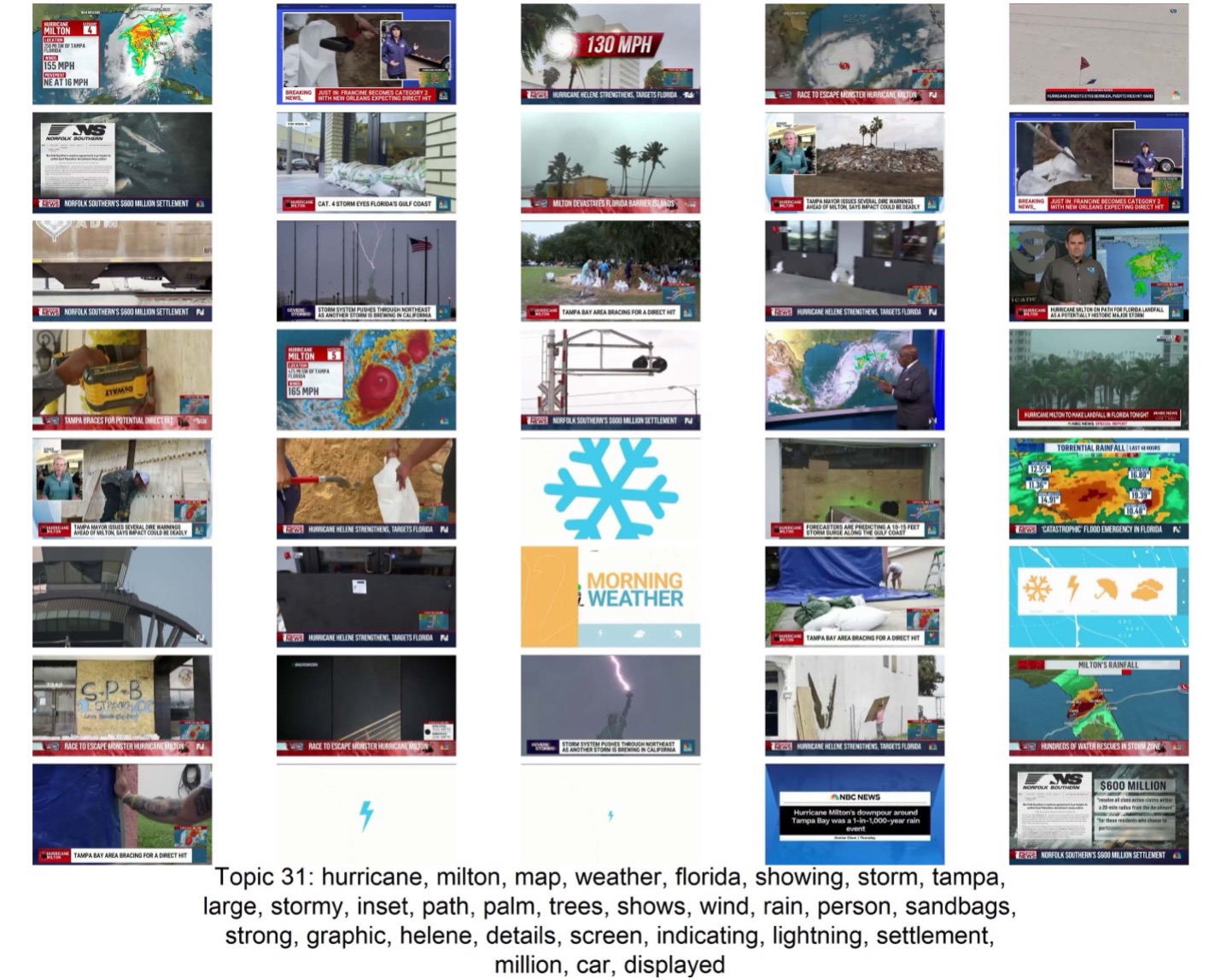} 
  \caption{Topic 31 Clustering Results.}
  \label{fig:figb31}
\end{figure}

\begin{figure}
  \centering
  \includegraphics[width=0.7\textwidth]{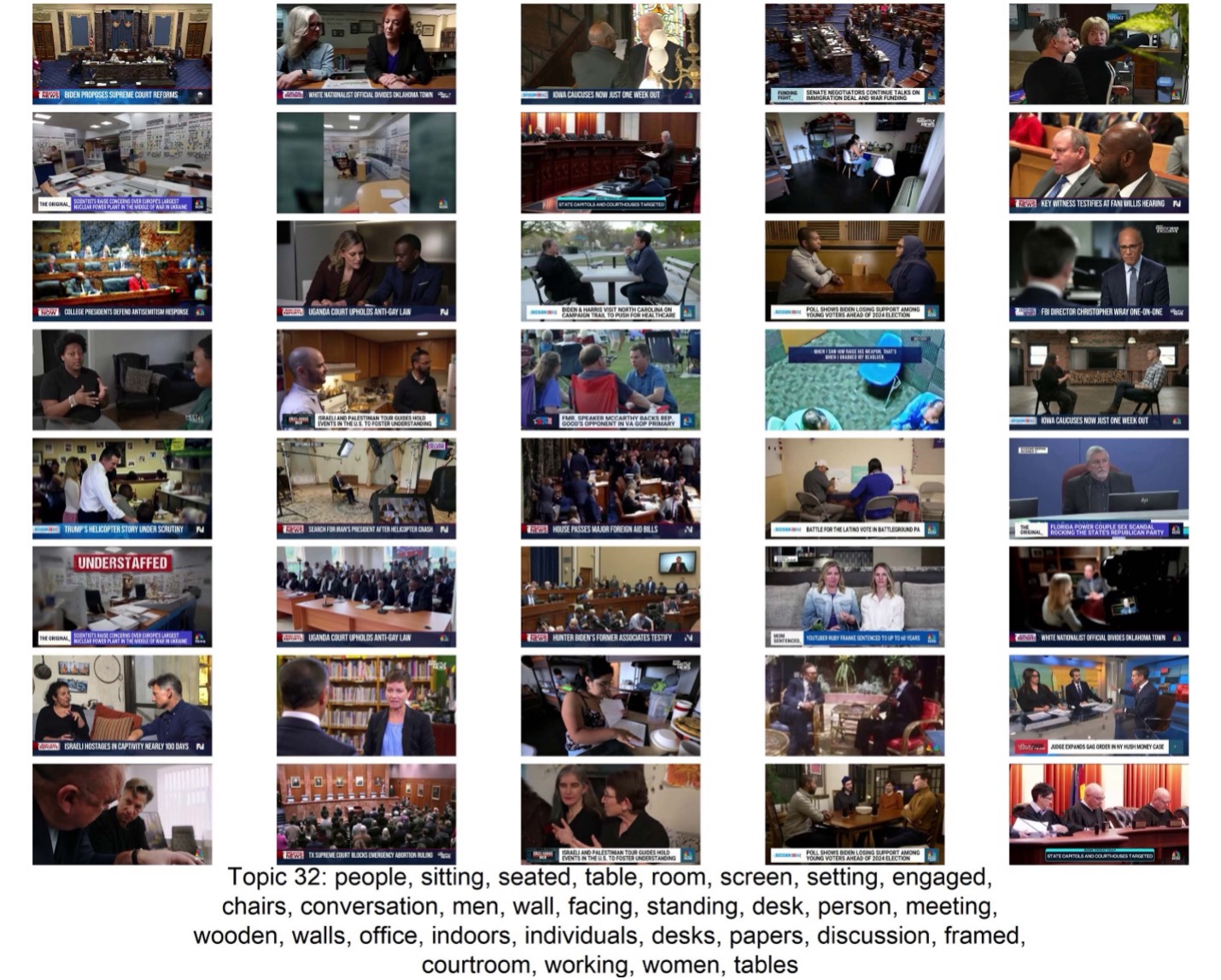} 
  \caption{Topic 32 Clustering Results.}
  \label{fig:figb32}
\end{figure}

\clearpage
\begin{figure}
  \centering
  \includegraphics[width=0.7\textwidth]{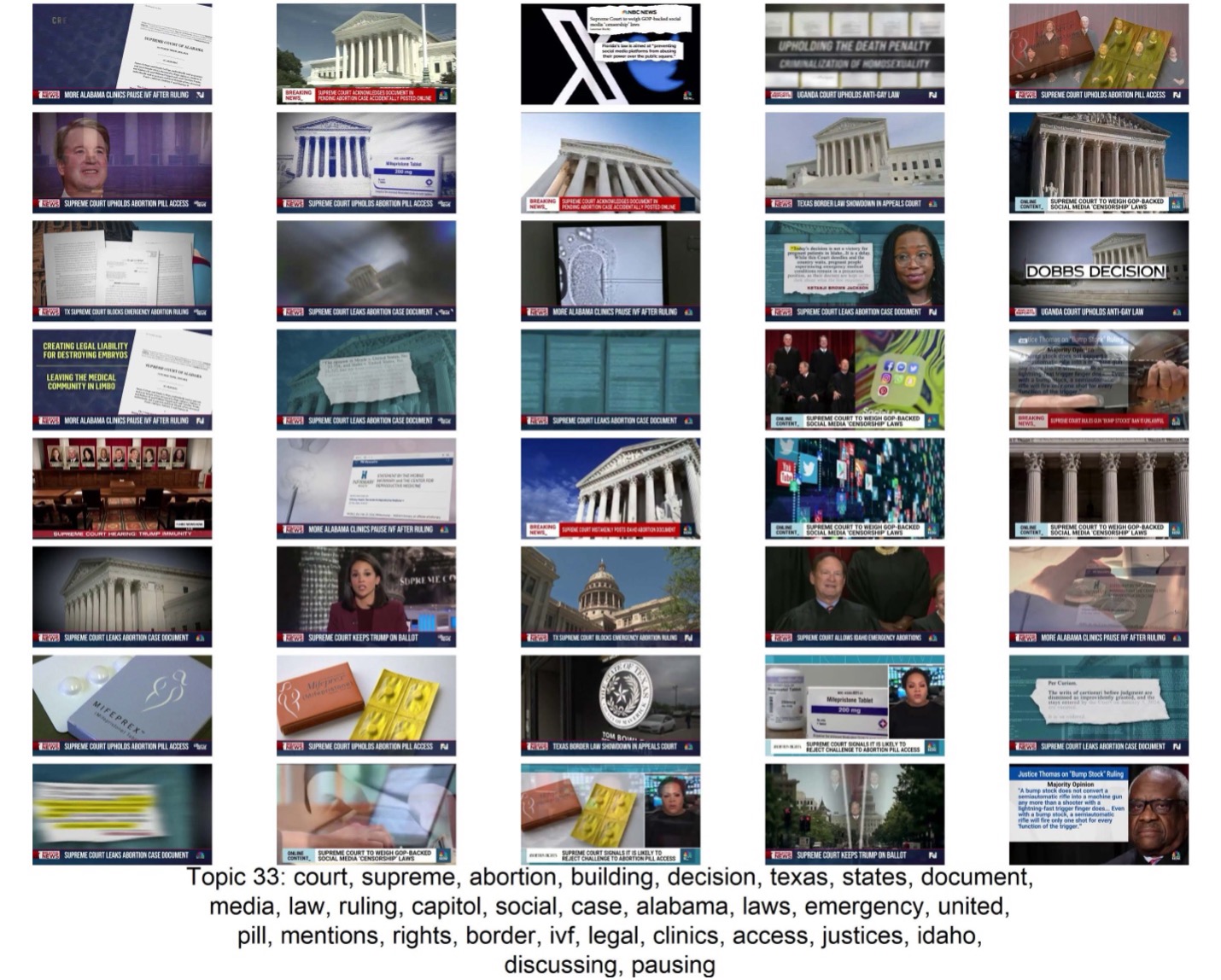} 
  \caption{Topic 33 Clustering Results.}
  \label{fig:figb33}
\end{figure}

\begin{figure}
  \centering
  \includegraphics[width=0.7\textwidth]{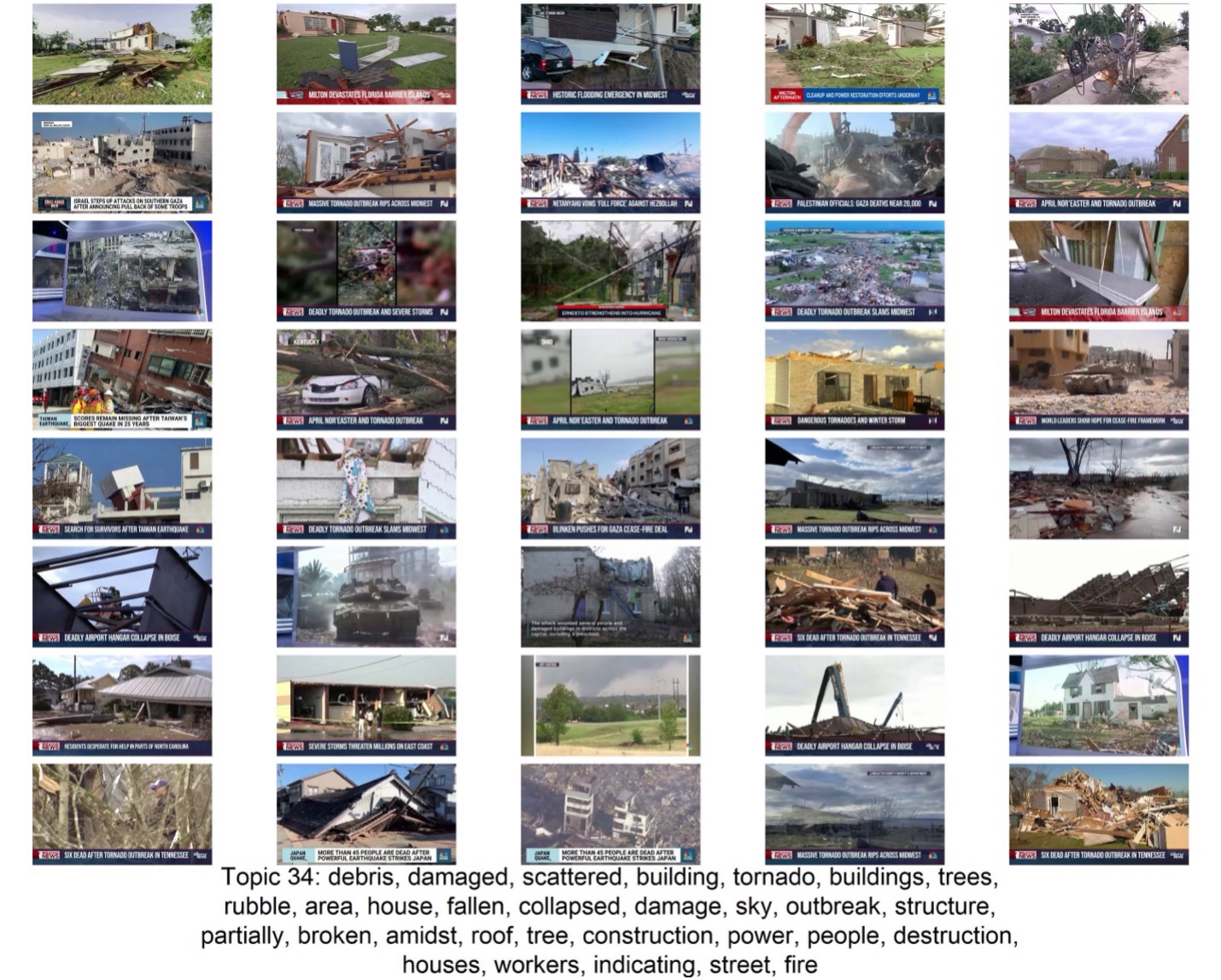} 
  \caption{Topic 34 Clustering Results.}
  \label{fig:figb34}
\end{figure}

\clearpage
\begin{figure}
  \centering
  \includegraphics[width=0.7\textwidth]{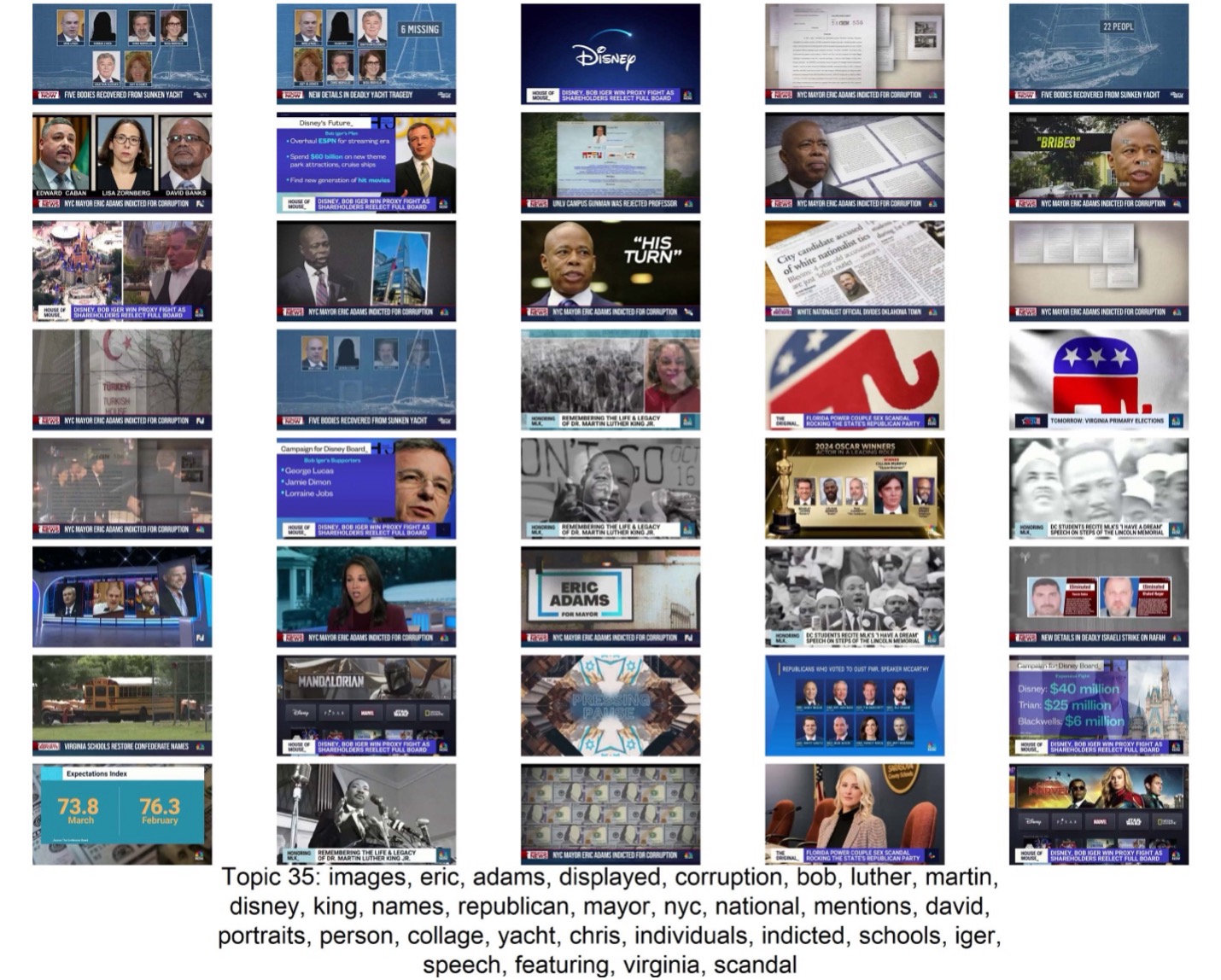} 
  \caption{Topic 35 Clustering Results.}
  \label{fig:figb35}
\end{figure}

\end{document}